\documentclass{aa}

\usepackage{graphicx}
\usepackage{txfonts}
\usepackage{natbib}
\bibpunct{(}{)}{;}{a}{}{,} 

\newcommand{\tothe}[1]{\times 10^{#1}}

\defcitealias{DiFolco07}{Paper~I}

\begin{document}

\title{A near-infrared interferometric survey of debris disc stars\thanks{Partly based on
observations collected at the European Southern Observatory, La Silla, Chile, under program IDs
073.C-0733, 077.C-0295 and 080.C-0712}}

\subtitle{II. CHARA/FLUOR observations of six early-type dwarfs}

\titlerunning{A near-infrared interferometric survey of debris disc stars. II}


\author{O.~Absil\inst{1}\fnmsep\thanks{Marie Curie EIF Postdoctoral Fellow.} \and
E.~Di~Folco\inst{2} \and A.~M\'erand\inst{3} \and J.-C.~Augereau\inst{1} \and
V.~Coud\'e~du~Foresto\inst{4} \and D.~Defr\`ere\inst{5} \and P.~Kervella\inst{4} \and
J.~P.~Aufdenberg\inst{6} \and M.~Desort\inst{1} \and D.~Ehrenreich\inst{1} \and
A.-M.~Lagrange\inst{1} \and G.~Montagnier\inst{2,1} \and J.~Olofsson\inst{1} \and
\\ T.~A.~ten~Brummelaar\inst{3} \and H.~A.~McAlister\inst{3} \and J.~Sturmann\inst{3} \and
L.~Sturmann\inst{3} \and N.~H.~Turner\inst{3}}

\offprints{O.~Absil}

\institute{LAOG--UMR 5571, CNRS and Universit\'e Joseph Fourier, BP 53, F-38041 Grenoble, France
 \\ \email{olivier.absil@obs.ujf-grenoble.fr}
 \and Observatoire Astronomique de l'Universit\'e de Gen\`eve, 51 chemin des Maillettes, CH-1290
      Sauverny, Switzerland
 \and Center for High Angular Resolution Astronomy, Georgia State University, PO Box 3969, Atlanta,
      Georgia 30302-3965, USA
 \and LESIA--UMR 8109, CNRS and Observatoire de Paris-Meudon, 5 place J.~Janssen, F-92195 Meudon,
      France
 \and Institut d'Astrophysique et de G\'eophysique, Universit\'e de Li\`ege, 17 All\'ee du Six
 Ao\^ut, B-4000 Li\`ege, Belgium
 \and Physical Sciences Department, Embry-Riddle Aeronautical University, Daytona Beach, FL 32114,
      USA}

\date{Received 21 April 2008; accepted 19 June 2008}

\abstract
{}
{We aim at directly detecting the presence of optically thin circumstellar dust emission within the
terrestrial planetary zone around main sequence stars known to harbour cold debris discs. The
present study focuses on a sample of six bright A- and early F-type stars.}
{High-precision interferometric observations have been obtained in the near-infrared $K$ band with
the FLUOR instrument installed on the CHARA Array. The measured squared visibilities are compared
to the expected visibility of the stellar photospheres based on theoretical photospheric models
taking into account rotational distortion. We search for potential visibility reduction at short
baselines, a direct piece of evidence for resolved circumstellar emission.}
{Our observations bring to light the presence of resolved circumstellar emission around one of the
six target stars ($\zeta$\,Aql) at the $5\sigma$ level. The morphology of the emission source
cannot be directly constrained because of the sparse spatial frequency sampling of our
interferometric data. Using complementary adaptive optics observations and radial velocity
measurements, we find that the presence of a low-mass companion is a likely origin for the excess
emission. The potential companion is characterised by a $K$-band contrast of four magnitudes. It
has a most probable mass of about $0.6 M_{\odot}$ and is expected to orbit between about 5.5\,AU
and 8\,AU from its host star assuming a purely circular orbit. Nevertheless, by adjusting a
physical debris disc model to the observed Spectral Energy Distribution of the $\zeta$\,Aql system,
we also show that the presence of hot dust within 10\,AU from $\zeta$\,Aql, producing a total
thermal emission equal to $1.69\pm 0.31$\% of the photospheric flux in the $K$ band, is another
viable explanation for the observed near-infrared excess. Our re-interpretation of archival near-
to far-infrared photometric measurements shows however that cold dust is not present around
$\zeta$\,Aql at the sensitivity limit of the IRS and MIPS instruments onboard Spitzer, and urges us
to remove $\zeta$\,Aql from the category of bona fide debris disc stars.}
{The hot debris disc around Vega (Absil et al.\ 2006) currently remains our only secure resolved
detection within the context of this survey, with six genuine early-type debris disc stars observed
so far. Further observations will be needed to assess whether $\zeta$\,Aql also belongs to this hot
debris disc category.}


\maketitle


\section{Introduction}

Debris discs are optically thin, gas-poor dust discs around main sequence (MS) stars. The presence
of circumstellar dust around stars with ages above $\sim$10~Myr is attributed to populations of
planetesimals that were neither used to make up planets nor ejected from the system by the time the
nebular gas was dispersed \citep{Mann06}. These leftovers produce dust by mutual collisions and
comet-type activity. Being continuously replenished by small bodies, the disc can then persist over
much of the star's lifetime. Due to its large total cross-section area, dust is much easier to
observe than planets, not to speak of planetesimals. On the other hand, distributions of dust
respond to the presence of planetary perturbers, reflect distributions of the parent bodies and
bear important memory of the planetary formation process in the past. Hence debris discs can be
used as sensitive tracers of planets, as well as small body populations, and should reflect
evolutionary stages of planetary systems. This explains the substantial effort invested in the
observation and modelling of debris discs over the last two decades.

Early-type stars, and A-type stars in particular, have been the most successfully studied targets
in the context of debris disc studies so far. This is due to a great extent to their intrinsic
brightness, which efficiently lights up their debris discs up to large distances (Earth-like
temperatures occur at about 5\,AU from such stars). Additionally, their MS lifetimes
($\sim$800~Myr) are long enough to encompass the main evolutionary stages of a typical planetary
system. Far-infrared surveys carried out with space satellites have been particularly successful in
detecting cold dust around nearby A-type stars. For instance, Spiter/MIPS observations of
160~A-type MS stars have shown that about 33\% of such stars possess significant excess emission at
24 and 70~$\mu$m \citep{Su06}, which represents a considerably higher excess rate than what has
been found for old solar analogs and M dwarfs \citep{Bryden06}.

Unlike these successful detections of cold debris in the far-infrared, the discovery of warm
($\sim$300\,K) or hot ($\sim$1000\,K) dust through mid- and near-infrared observations has been
limited to a very small number of targets so far. Furthermore, the rare detections have generally
been obtained for young MS stars showing strong mid-infrared silicate features, such as $\beta$~Pic
\citep[$\sim$12\,Myr,][]{Pantin97}, HD\,145263 \citep[$\sim$8\,Myr,][]{Honda04}, $\eta$~Tel
\citep[$\sim$12\,Myr,][]{Chen06} or HD\,172555 \citep[$\sim$12\,Myr,][]{Chen06}. At such ages,
planetary systems are still expected to be in the process of forming planets, especially in the
inner part of the disc ($<10$\,AU) where rocky planets may take up to one hundred Myr to accrete
most of the small bodies on nearby orbits and reach their final mass \citep[see][for a solar-mass
star]{Kenyon06}. The observed large mid-infrared excesses are therefore expected to be related to
the end of the planet building phase rather than to the dust produced by the collisional grinding
of an evolved planetary system, similar to the zodiacal dust in our solar system. For ``mature''
A-type stars, the lack of near- to mid-infrared excess has generally been associated with a dearth
of warm dust grains, a suggestion generally confirmed by the presence of inner holes in resolved
images around famous objects such as $\alpha$~PsA~\citep{Kalas05} or $\alpha$~Lyr~\citep{Su05}.

Incidentally, the inner part of debris discs is a very interesting region to study, as it directly
probes the location where planets are supposed to have formed and evolved---conversely, the outer
part of the disc ($>20$\,AU), similarly to the Kuiper belt in our solar system, only bears a memory
of the outermost massive planets through gravitational interactions such as mean motion resonances
\citep[see e.g.][]{Reche08}. The characterisation of inner debris discs is therefore crucial for an
understanding of the formation, evolution and dynamics of planetary systems (including our own
solar system), as well as to set the scene for the emergence of life on rocky planets.

Reaching a better sensitivity to the inner part of debris discs has thus been an important
challenge during the past years, and the detection of small amounts of hot dust around mature
early-type stars has finally been enabled by the advent of high-precision near-infrared stellar
interferometry. Using the VLTI/VINCI instrument, \citet{DiFolco04} derived upper limits to the
$K$-band dust emission around five stars. The first robust resolved detection of hot dust was
obtained for the bright A0V-type star $\alpha$~Lyr (Vega) by \citet{Absil06} with CHARA/FLUOR on an
optimised set of interferometric baselines, showing the presence of extended emission accounting
for $1.29 \pm 0.19$\% of the photospheric flux in the $K$ band. This result was already suggested
by \citet{Ciardi01}, although with a large uncertainty on the flux ratio.

This recent detection has raised a number of questions regarding the nature and origin of inner
dust grains. In particular, a scenario involving the presence of star-grazing comets, injected into
the inner planetary system by dynamical perturbations caused by migrating planets, has been
proposed, inspired by the Late Heavy Bombardment (LHB) that happened in the early solar
system~\citep{Gomes05}. A similar scenario is also proposed by \citet{Wyatt07a} to explain the
presence of warm dust ($\sim$300\,K) detected by the Spitzer Space Telescope around a few
solar-type stars. We have thus decided to initiate a near-infrared interferometric survey of nearby
debris disc stars to assess the occurrence of hot excesses around MS stars. The first paper of this
series \citep[][hereafter \citetalias{DiFolco07}]{DiFolco07} focused on two solar-type stars,
showing the presence of hot dust around $\tau$\,Cet. In this paper, we discuss early-type stars,
which hold a privileged position in our survey because their brightness makes them well suited for
an interferometric study.


\section{Methodology and stellar sample}

The principle of debris disc detection by stellar interferometry is based on the fact that the
stellar photosphere and its surrounding dust disc have different spatial scales. For an A-type star
at a distance of 20\,pc, the angular diameter of the photosphere is typically about 1\,mas, while
the circumstellar disc extends beyond the sublimation radius of dust grains, typically located
around 10 to 20\,mas for black body grains sublimating at $T_{\rm sub}\simeq1500$\,K. In the
infrared $K$ band, the debris disc is therefore generally fully resolved at short baselines
($10-20$\,m), while the photosphere is only resolved at long baselines ($\sim$200\,m). One can take
advantage of this fact to isolate the contribution of circumstellar dust by performing visibility
measurements at short baselines, where the stellar photosphere is almost unresolved. The presence
of resolved circumstellar emission then shows up as a deficit of squared visibility with respect to
the expected visibility of the bare stellar photosphere, as shown in \citetalias{DiFolco07}:
\begin{equation}
{\cal V}^2(b) \simeq (1-2\epsilon_{\rm CSE}) {\cal V}_{\ast}^2(b) \, ,
\end{equation}
where ${\cal V}^2$ and ${\cal V}_{\ast}^2$ are respectively the squared visibility of the star-disc
system and of the bare stellar photosphere, $b$ is the interferometer baseline length, and
$\epsilon_{\rm CSE}$ is the flux ratio between the integrated circumstellar emission within the
field-of-view and the stellar photospheric emission. This equation is valid only for short
baselines and for $\epsilon_{\rm CSE} \ll 1$, as expected for an optically and geometrically thin
circumstellar disc.

Due to the expected faintness of the circumstellar emission ($\epsilon_{\rm CSE} \lesssim 1$\%),
detection needs both a high accuracy on the measured ${\cal V}^2$ and on the estimation of the
squared visibility ${\cal V}_{\ast}^2$ of the bare photosphere at short baselines. While the former
has already been demonstrated in the framework of single-mode near-infrared interferometers
\citep[e.g.,][]{Kervella03}, the latter is ensured to a large extent by the fact that the stellar
photosphere is almost unresolved at short baselines, so that we can tolerate a certain level of
imprecision on the knowledge of the stellar angular diameter without jeopardising the detection.
For instance, a stellar photosphere with an angular diameter of $1.0 \pm 0.1$\,mas produces a
squared visibility of $0.995 \pm 0.001$ at a baseline of 20\,m in the $K$ band. Therefore, in the
following study, diameter measurements at long interferometric baselines are not required to derive
accurate photospheric models, and empirical surface-brightness relations \citep{Kervella04} can
instead be used to predict stellar angular diameters with a sufficient accuracy, taking into
account rotational distortion.

To start our survey of early-type debris disc stars, we have chosen six bright ($K<4$) nearby
($\sim$20\,pc) dwarfs, with spectral types ranging from A0\,V to F2\,V. The main characteristics of
our six target stars are summarised in Table~\ref{tab:diam}, together with the already-discussed
$\alpha$\,Lyr \citep{Absil06}. All of them have been classified as debris disc stars based on the
measurement of a mid- or far-infrared excess emission above the expected photospheric level, using
infrared satellites such as IRAS, ISO or the Spitzer Space Telescope.


\section{Observations and data reduction}

Interferometric observations were obtained in the infrared $K$ band ($1.94 - 2.34$~$\mu$m) with
FLUOR, the Fiber Linked Unit for Optical Recombination \citep{Coude03}, using the shortest baseline
of the CHARA Array formed by the S1 and S2 telescopes \citep[34~metres long,][]{tenBrummelaar05}.
Observations took place during Spring 2006, between April 30th and May 11th.

\begin{table*}[!t]
\caption{Fundamental parameters and estimated angular diameters for the seven selected targets
(including $\alpha$~Lyr). A mean limb-darkened diameter is first computed using
Eq.~\ref{eq:diam}---this actually represents the geometric mean of the minor and major axes of the
elliptical photosphere. The apparent photospheric oblateness is then estimated by means of
Eq.~\ref{eq:oblateness}, from which we finally deduce the angular diameters for the apparent minor
and major axes of the photosphere. The 1$\sigma$ errors are given in superscript.} \label{tab:diam}
\begin{center}
\begin{tabular}{ccccccccccccc}
\hline \hline Name & HD & Type & Dist. & Mass & $v \sin i$ & $m_V$ & $m_K$ & Mean $\theta_{\rm LD}$
& Appar. & Minor $\theta_{\rm LD}$ & Major $\theta_{\rm LD}$ & Refs.
\\ & & & (pc) & ($M_{\odot}$) & (km\,s$^{-1}$) &  &  & (mas) & oblat. & (mas) & (mas) &
\\ \hline
\object{$\beta$~UMa}  &  95418 & A1\,V & 24.3 & 2.28 & 46 & $2.35^{0.01}$ & $2.38^{0.06}$ &
$1.095^{0.022}$ & 1.007 & $1.091^{0.022}$ & $1.099^{0.022}$ & 1, 4, 6, 7 \\
\object{$\eta$~Crv}   & 109085 & F2\,V & 18.2 & 1.44 & 92 & $4.30^{0.01}$ & $3.54^{0.05}$ &
$0.736^{0.015}$ & 1.022 & $0.728^{0.015}$ & $0.744^{0.015}$ & 1, 4, 6, 8 \\
\object{$\sigma$~Boo} & 128167 & F2\,V & 15.5 & 1.28 & 15 & $4.47^{0.01}$ & $3.49^{0.02}$ &
$0.783^{0.016}$ & 1.001 & $0.783^{0.016}$ & $0.783^{0.016}$ & 1, 4, 6, 9 \\
\object{$\alpha$~CrB} & 139006 & A0\,V & 22.9 & 2.58 & 138 & $2.23^{0.01}$ & $2.20^{0.05}$ &
$1.202^{0.024}$ & 1.059 & $1.168^{0.024}$ & $1.237^{0.024}$ & 2, 4, 6, 10 \\
\object{$\gamma$~Oph} & 161868 & A0\,V & 29.1 & 2.18 & 210 & $3.75^{0.01}$ & $3.67^{0.05}$ &
$0.616^{0.012}$ & 1.107 & $0.585^{0.012}$ & $0.648^{0.012}$ & 1, 4, 6, 11 \\
\object{$\alpha$~Lyr} & 172167 & A0\,V &  7.8 & 2.30 &  22 & $0.03^{0.01}$ & $0.00^{0.02}$ &
$3.312^{0.067}$ & 1.002 & $3.309^{0.067}$ & $3.315^{0.067}$ & 3, 5, 6, 12 \\
\object{$\zeta$~Aql}  & 177724 & A0\,V & 25.5 & 2.37 & 317 & $2.99^{0.01}$ & $2.90^{0.02}$ &
$0.880^{0.018}$ & 1.307 & $0.770^{0.018}$ & $1.006^{0.018}$ & 1, 4, 6, 13 \\
\hline
\end{tabular}
\end{center}
References.---Masses from (1) \citealt{AllendePrieto99}, (2) \citealt{Tomkin86}, (3)
\citealt{Aufdenberg06}; $v \sin i$ from (4) \citealt{Royer02}, (5) \citealt{Hill04}; $V$ magnitudes
from (6) \citealt{Perryman97}; $K$ magnitudes from (7) \citealt{Neugebauer97}, (8)
\citealt{Sylvester96}, (9) \citealt{Blackwell79}, (10) \citealt{Sneden78}, (11) \citealt{Selby88},
(12) \citealt{Aumann91}, (13) \citealt{Leggett86}.

\end{table*}

The FLUOR field-of-view, limited by the use of single-mode fibres, has a Gaussian shape resulting
from the overlap integral of the incoming turbulent wave fronts with the fundamental mode of the
fiber \citep{Guyon02}. Its actual size depends on the atmospheric turbulence conditions. The
estimation of the mean Fried parameter $\langle r_0 \rangle$ provided by the CHARA array during our
observations (7~cm in the visible) allows us to derive a mean value of $0\farcs8$ for the full
width at half maximum of the field-of-view. This parameter is sensitive to the actual atmospheric
conditions for each individual target star (it typically ranges between $0\farcs7$ and $1\farcs2$),
but the only associated effect is to widen the search region for circumstellar emission.

The FLUOR Data Reduction Software \citep[DRS,][]{Coude97,Kervella04,Merand06} was used to extract
the raw squared modulus of the coherence factor between the two independent apertures. The
extraction of the squared visibilities from the fringe packets recorded in the time domain is based
on the integration of the squared fringe peak obtained by a Fourier transform of the fringe packet
\citep{Coude97}, or equivalently by a wavelet analysis \citep{Kervella04}. The interferometric
transfer function of the instrument was estimated by observing calibrator stars before and after
each observation of a scientific target. All calibrator stars (listed in Table~\ref{tab:calib})
were chosen from two catalogues developed for this specific purpose \citep{Borde02,Merand05}.
Calibrators chosen in this study are late G or K giants, whereas our target stars have spectral
types between A0 and F2. Since the visibility estimator implemented in the FLUOR DRS depends on the
actual spectrum of the target star, an appropriate correction must be applied to our data,
otherwise our squared visibilities would be biased at a level of about 0.3\% \citep{Coude97}. This
correction can be based either on the shape factors discussed by \citet{Coude97} or on a wide band
model for estimating the calibrator's visibilities and interpreting the data \citep[see
e.g.][]{Kervella03,Aufdenberg06}. The latter method was chosen for this work, and all the
calculations presented here therefore take into account a full model of the FLUOR instrument,
including the spectral bandwidth effects.

\begin{table}[t]
\caption{Calibrators chosen from the catalogues of \citet{Borde02} and \citet{Merand05}, listed
with spectral type, $K$ magnitude, uniform disc (UD) angular diameter in $K$ band with $1\sigma$
error bar, and associated scientific target(s).} \label{tab:calib} \centering
\begin{tabular}{ccccc}
\hline \hline   Identifier    & Sp.~type & $m_K$ & $\theta_{\rm UD}$ (mas) & Targets
\\ \hline \object{46\,Boo}    & K2\,III  & 2.66  & $1.433 \pm 0.019$    & $\alpha$~CrB
\\        \object{9\,CrB}     & G9\,III  & 3.07  & $1.101 \pm 0.014$    & $\alpha$~CrB
\\        \object{QY\,Ser}    & K8\,IIIb & 1.53  & $2.680 \pm 0.029$    & $\alpha$~CrB
\\        \object{HD\,92095}  & K3\,III  & 2.20  & $1.527 \pm 0.020$    & $\beta$~UMa
\\        \object{HD~100615}  & K0\,III  & 3.19  & $1.027 \pm 0.014$    & $\beta$~UMa
\\        \object{44\,UMa}    & K3\,III  & 2.04  & $1.990 \pm 0.023$    & $\beta$~UMa
\\        \object{$\chi$~Vir} & K2\,III  & 2.02  & $1.960 \pm 0.023$    & $\eta$~Crv
\\        \object{HD\,108522} & K4\,III  & 3.46  & $1.171 \pm 0.016$    & $\eta$~Crv
\\        \object{HD\,111500} & K4\,III  & 3.05  & $1.395 \pm 0.020$    & $\eta$~Crv
\\        \object{HD\,157617} & K1\,III  & 2.71  & $1.305 \pm 0.018$    & $\gamma$~Oph
\\        \object{71\,Oph}    & G8\,III  & 2.43  & $1.500 \pm 0.020$    & $\gamma$~Oph,
\\                            &          &       &                   & $\sigma$~Boo, $\zeta$~Aql
\\        \object{HD\,162113} & K0\,III  & 3.62  & $0.904 \pm 0.012$    & $\gamma$~Oph
\\        \object{HD\,162468} & K1\,III-IV& 3.17 & $1.123 \pm 0.015$    & $\gamma$~Oph
\\        \object{HD\,133392} & G8\,III  & 3.26  & $1.052 \pm 0.014$    & $\sigma$~Boo
\\        \object{HD\,132304} & K3\,III  & 3.48  & $1.042 \pm 0.014$    & $\sigma$~Boo
\\        \object{HD\,126597} & K2\,III  & 3.46  & $1.004 \pm 0.014$ & $\sigma$~Boo, $\gamma$~Oph
\\        \object{HD\,176527} & K2\,III  & 2.04  & $1.721 \pm 0.024$    & $\zeta$~Aql
\\        \object{HD\,175743} & K1\,III  & 3.35  & $1.106 \pm 0.015$    & $\zeta$~Aql
\\        \object{$\xi$~Aql} & G9.5\,IIIb& 2.37  & $1.620 \pm 0.021$    & $\zeta$~Aql
\\        \object{$\mu$~Aql} & K3\,IIIb  & 1.76  & $2.240 \pm 0.023$    & $\zeta$~Aql
\\ \hline
\end{tabular}
\end{table}


\section{Estimating stellar angular diameters}

A reliable estimation of the photospheric angular diameter is an important pre-requisite for the
detection of circumstellar emission at short baselines. In this section, we use surface-brightness
relations to estimate the mean photospheric diameter of our target stars. We then apply a standard
model of a rotating photosphere in hydrostatic equilibrium to estimate the rotation-induced
distortion. Our approach is finally validated by comparison with long-baseline interferometric
measurements where available.

    \subsection{Diameter estimation from surface-brightness relations} \label{sub:diam}

Accurate angular diameter estimations for MS stars can be obtained by applying empiric
surface-brightness relations \citep[see e.g.,][]{Kervella04,DiBenedetto05}, using as input the
measured magnitudes in two standard photometric bands. These relations have been calibrated by
interferometric measurements \citep{Kervella04}, and are valid for spectral types from A0 to M2.
The smallest intrinsic dispersion ($\sigma \le 1$\%) is obtained by combining $K$- or $L$-band
magnitudes with visible magnitudes ($B$ or $V$). Here, we choose to use the $K$-band magnitudes,
which are generally known with good accuracy, and combine them with the $V$-band magnitudes
measured by {\sc Hipparcos} \citep{Perryman97}. $K$ magnitudes have been preferred to $L$
magnitudes, because the potential presence of hot dust around the target stars is expected to
affect the $K$ band at a smaller level than the $L$ band---the contrast between a $\sim$9000\,K
photosphere and $\sim$1000\,K surrounding dust decreases with increasing wavelength. In practice,
we do not expect the $K$-band magnitudes to be affected by dust emission by an amount larger than
5\%, as for higher contributions the signature of hot dust would have already been detected. Our
interferometric measurements will confirm this assumption {\em a posteriori}. The ($V$, $V-K$)
relation, where the $V$ magnitude is used to calibrate the reference magnitude, has been chosen in
the present study because of its robustness to dust emission and because the accuracy of the $V$
magnitudes \citep[0.01~mag\footnote{This estimated error bar is conservative (or even pessimistic)
for the bright stars that we are studying here \citep{vanLeeuwen97}.},][]{Perryman97} is better
than that of the $K$ magnitudes. The final error bar on the estimated limb-darkened angular
diameter $\theta_{\rm LD}$ is therefore generally smaller using the ($V$, $V-K$) relation than the
($K$, $V-K$) relation. The mathematical expression of the ($V$, $V-K$) relation is \citep[see
Table~4 from][]{Kervella04}:
\begin{equation}
\log \theta_{\rm LD}(V, V-K) = 0.2753 (V-K) + 0.5175 - 0.2V \, . \label{eq:diam}
\end{equation}
It has a dispersion of 1\% and shows no significant non linearity. The angular diameter estimations
for our seven targets are given in the ninth column of Table~\ref{tab:diam} (``Mean $\theta_{\rm
LD}$'').

The associated limb darkening coefficients $u_K$ can be found in \citet{Claret95}, assuming a
linear limb-darkening law $I_{\lambda}(\mu) = 1-u_{\lambda}(1-\mu)$ as in \citet{Kervella04}, where
$\mu$ is the cosine of the angle between the perpendicular to the surface of the star and the line
of sight. The closest tabulated models to the physical properties of the stars were chosen. It must
be noted that the rapid rotation of some of our targets may lead to substantial gravity darkening,
so that the tabulated limb darkening coefficients could be underestimated. Fortunately, considering
a photosphere 1.0\,mas in diameter observed with a 30\,m baseline, even a threefold increase of the
limb darkening coefficient would produce a variation smaller than 0.1\% on the squared visibility
(see Eq.~\ref{eq:visld} in Sect.~\ref{sec:circum}). The case of $\alpha$~Lyr, which is slightly
resolved on a 30-m baseline, must however be considered with additional care (see
Sect.~\ref{sub:diaminterf}).

Estimating the angular diameter of $\alpha$~CrB is a little more complicated than for the other six
targets, as it is actually a binary system of Algol type. The primary and secondary are estimated
to be A0\,V and G5\,V stars respectively, and the magnitude difference in $V$ band is estimated to
be $\Delta M_V=4.89$ \citep{Tomkin86}. Based on a $V$-band magnitude of $2.22\pm0.01$ for the
system \citep{Perryman97}, we adopt a $V$-band magnitude of $2.23\pm0.01$ for the primary. Then,
using the intrinsic colours of MS stars in the infrared \citep{Bessel88}, we compute the magnitude
difference between the primary and secondary in the $K$ band ($\Delta M_K=3.25$), and derive the
expected $K$-band photometry of the primary from the system's photometry in a similar way. The
original $K$ magnitude for the binary system, $K=2.14\pm0.04$ \citep{Sneden78}, gives a corrected
magnitude of $2.20\pm0.05$, where the error bar has been slightly revised to account for possible
errors in the estimation of the spectral types and intrinsic colours of the two components.

    \subsection{Effect of rapid rotation}

With projected rotational velocities up to 317\,km\,s$^{-1}$ (Table~\ref{tab:diam}), we expect the
photospheres of our target stars to deviate significantly from spherical symmetry due to large
centrifugal forces. We will assume in the following study that the apparent photospheric shape can
be modelled by an ellipse with semi-major and semi-minor axes noted $R_a$ and $R_b$, and we force
the angular surface of the ellipse to be equal to the angular surface derived from
surface-brightness relations in the previous section:
\begin{equation}
R_a R_b = R_m^2 = \frac{\theta_{\rm LD}^2d^2}{4} \; ,
\end{equation}
where $R_m$ is the radius of the equivalent spherical photosphere (i.e., the geometric mean of the
major and minor radii) and $d$ the distance to the target star. Under the assumptions of
hydrostatic equilibrium, uniform rotation, and of a point mass gravitational potential, the
apparent rotational velocity $v \sin i$ can be related to the major and minor radii $R_a$ and $R_b$
with the following equation \citep{Tassoul78}:
\begin{equation}
v \sin i \simeq \sqrt{2GM \left( \frac{1}{R_b}-\frac{1}{R_a} \right)} \, .
\end{equation}
We define the {\em apparent oblateness} ($\rho$) of the photosphere as the ratio of the major and
minor radii ($\rho = R_a/R_b$), and solve the above equation for $\rho$ as a function of $R_m$ to
obtain:
\begin{equation}
\rho \simeq \left( \frac{(v \sin i)^2 R_m}{4GM} + \sqrt{1 + \left( \frac{(v \sin i)^2 R_m}{4GM}
\right)^2} \right)^2 \, . \label{eq:oblateness}
\end{equation}
Using the limb-darkened angular radii, masses, and $v \sin i$ listed in Table~\ref{tab:diam}, we
derive the apparent oblateness of the photospheres of our seven target stars. They range between
1.001 and 1.307 (see Table~\ref{tab:diam}). From these estimations of oblateness, we can derive the
expected angular sizes of the major and minor axes. Since the orientations of the photospheres are
unknown, the differences between the major and minor angular radii directly give the additional
uncertainty on the stellar diameters along the CHARA S1--S2 baseline due to stellar oblateness. The
additional error is typically a few percent of the mean stellar diameter, and reaches 14\% for the
fastest rotator.

    \subsection{Long-baseline interferometric measurements} \label{sub:diaminterf}

\begin{table*}[!t]
\caption{Application of surface-brightness relations and of our standard rotating photosphere model
to four rapidly rotating stars that have been directly measured with infrared stellar
interferometers. The agreement between our estimated mean limb-darkened stellar diameter
(``$\theta_{\rm LD}$ model'') and oblateness (``$\rho$ model'') with the measured stellar
parameters is very satisfactory. The largest discrepancy between model and observations is found
for $\alpha$\,Aql and is less than 3\%---it should be noted that the interferometric measurements
reported on this particular target by \citet{Monnier07} and \citet{Domiciano05} disagree at the
$3\sigma$ level (4\% of $\theta_{\rm LD}$).} \label{tab:check}
\begin{center}
\begin{tabular}{ccccccccc}
\hline \hline Name & Type &$v \sin i$&Mass&$\theta_{\rm LD}$&$\theta_{\rm LD}$&$\rho$&$\rho$&Ref.
\\                   & & (km\,s$^{-1}$) & ($M_{\odot}$) & model & meas. & model & meas. &
\\ \hline
\object{$\alpha$~Leo} & B7\,V  & 317 & 3.4 & 1.47 & 1.47 & 1.323 & 1.320 & (1) \\
\object{$\alpha$~Lyr} & A0\,V  & 21.9& 2.3 & 3.31 & 3.31 & 1.0015 & 1.0014 & (2,3) \\
\object{$\alpha$~Cep} &A7\,IV-V& 283 & 2.0 & 1.51 & 1.54 & 1.296 & 1.298 & (4) \\
\object{$\alpha$~Aql} & A7\,V  & 240 & 1.8 & 3.31 & 3.39 & 1.170 & 1.164 & (5) \\
\hline
\end{tabular}
\end{center}
References: (1) \citet{McAlister05}, (2) \citet{Aufdenberg06}, (3) \citet{Peterson06b}, (4)
\citet{vanBelle06}, (5) \citet{Monnier07}.
\end{table*}

Since long-baseline infrared interferometric measurements are generally not available for our
target stars, we will try to validate our approach by comparison with interferometric measurements
of similar MS stars. For the first step of our modelling approach, i.e., surface-brightness
relations, such a validation has already been thoroughly addressed by \citet{Kervella04}. Now, for
the second step, i.e., the elliptic photospheric model, we can test the validity of
Eq.~\ref{eq:oblateness} on four rapidly rotating stars with similar masses to our target stars,
whose apparent oblateness has been derived from interferometric measurements (see
Table~\ref{tab:check}). The agreement between the modelled apparent oblateness (``$\rho$~model'')
and the observed oblateness (``$\rho$~meas.'') is very good.

To improve the accuracy of our angular diameter estimations, actual long baseline interferometric
measurements would however be valuable. In the case of $\alpha$~Lyr, such measurements are
available, and have been used to produce a realistic model of the stellar photosphere in the $K$
band \citep{Aufdenberg06}. Using the $K$-band limb profile of \citet{Aufdenberg06}, we have derived
an accurate limb-darkened model for the stellar photosphere, giving the following parameters:
angular diameter $\theta_{\rm LD}=3.305\pm0.010$\,mas and linear limb-darkening coefficient
$u_K=0.361$\footnote{The limb-darkening coefficient varies significantly across the $K$ band. We
have computed here a broad-band coefficient integrated on the FLUOR pass band.}. The angular
diameter is slightly smaller than the physical diameter (3.33\,mas) derived by \citet{Aufdenberg06}
because the outermost layers of the stellar atmosphere, from $0.992 R_{\ast}$ to $R_{\ast}$ (where
the pressure is less than 0.01 dynes/cm$^2$), does not contribute significantly to the flux
emission. The angular diameter obtained from surface-brightness relations ($3.312\pm0.067$\,mas,
see Table~\ref{tab:diam}) is in good agreement with this measurement, although it must be noted
that the associated error bar is 7 times larger than that of the directly measured diameter.
Nonetheless, we will see in Sect.~\ref{sec:circum} that the accuracy on the modelled angular
diameters displayed in Table~\ref{tab:diam} is generally sufficient to allow the detection of
circumstellar emission at a level as low as 0.1\%. The case of $\alpha$~Lyr must be considered
separately, because its 3.305\,mas photosphere is partly resolved on a 30\,m baseline in $K$ band.
The error on the modelled photospheric diameter would then induce a large enough uncertainty on the
stellar ${\cal V}^2$ (about 0.6\%) to compromise the detection of circumstellar emission. Moreover,
due to gravity darkening, the actual limb-darkening coefficient is much larger than the tabulated
one \citep[0.20 in][]{Claret95}. Therefore, in the rest of this study, we will use for $\alpha$~Lyr
the photospheric model of \citet{Aufdenberg06} described here above.

Another target in our sample has been observed with long-baseline interferometry, this time in the
visible range: $\zeta$~Aql was used as a reference star by \citet{Peterson06a} to calibrate
interferometric observations of Altair with the Navy Prototype Optical Interferometer (NPOI). Using
archival NPOI data, the authors derived a preliminary model for the $\zeta$~Aql photosphere, giving
a minimum angular diameter of $0.815\pm0.005$\,mas. Even though this value is well within our
angular diameter range (see Table~\ref{tab:diam}), it is $2.5\sigma$ from the minimum angular
diameter that we derive from our rapidly rotating photospheric model ($0.770\pm0.018$\,mas). This
discrepancy could partly be due to the wavelength difference between the observations ($V$ band)
and the model ($K$ band).

We also note that $\sigma$~Boo has been observed as a calibrator by \citet{Boden05} at the Palomar
Testbed Interferometer (PTI) and at the NPOI. The authors adopted an angular diameter of
$0.77\pm0.04$\,mas based on SED modelling of broadband photometry, a value which is consistent with
our estimation. They did not mention any significant deviation from that estimated value based on
their measurements.

Finally, the eclipsing binary nature of $\alpha$~CrB provides an additional check of our model:
\citet{Tomkin86} derive a limb-darkened radius of $3.04\pm0.03~R_{\odot}$ for the primary, which
translates into an angular diameter of $1.235\pm0.012$\,mas. This value is most probably
representative of the major axis of the photosphere as the rotation axis of the primary and the
orbital axis of the binary system are expected to be aligned. The agreement with our major angular
diameter ($1.237\pm0.024$\,mas) is very good.


\section{Searching for circumstellar emission} \label{sec:circum}

To assess the presence of circumstellar emission around a target star, we compare the
interferometric ${\cal V}^2$ measurements with the expected squared visibility of the stellar
photosphere, using the oblate photospheric model presented in the previous section. Assuming a
linear limb darkening law, the monochromatic stellar visibility is given by \citep{HanburyBrown74}:
\begin{equation}
{\cal V}_{\lambda,\,\ast}^2 = \left( \frac{1-u_{\lambda}}{2} - \frac{u_{\lambda}}{3} \right)^{-2}
\left( (1-u_{\lambda}) \frac{J_1(x)}{x} + \frac{u_{\lambda}\sqrt{\pi}}{\sqrt{2}}
\frac{J_{3/2}(x)}{x^{3/2}} \right)^2 \, , \label{eq:visld}
\end{equation}
where $x=\pi \theta_b b / \lambda$, where $b$ is the interferometer baseline length projected on
the sky plane and $\theta_b$ is the limb-darkened diameter along the direction of the baseline. The
wide-band visibilities are then computed by integrating the above expression over the FLUOR
bandpass, using the stellar spectrum as a weighting factor \citepalias[see][]{DiFolco07}.

\begin{figure*}[!t]
\centering
\resizebox{\hsize}{!}{\includegraphics{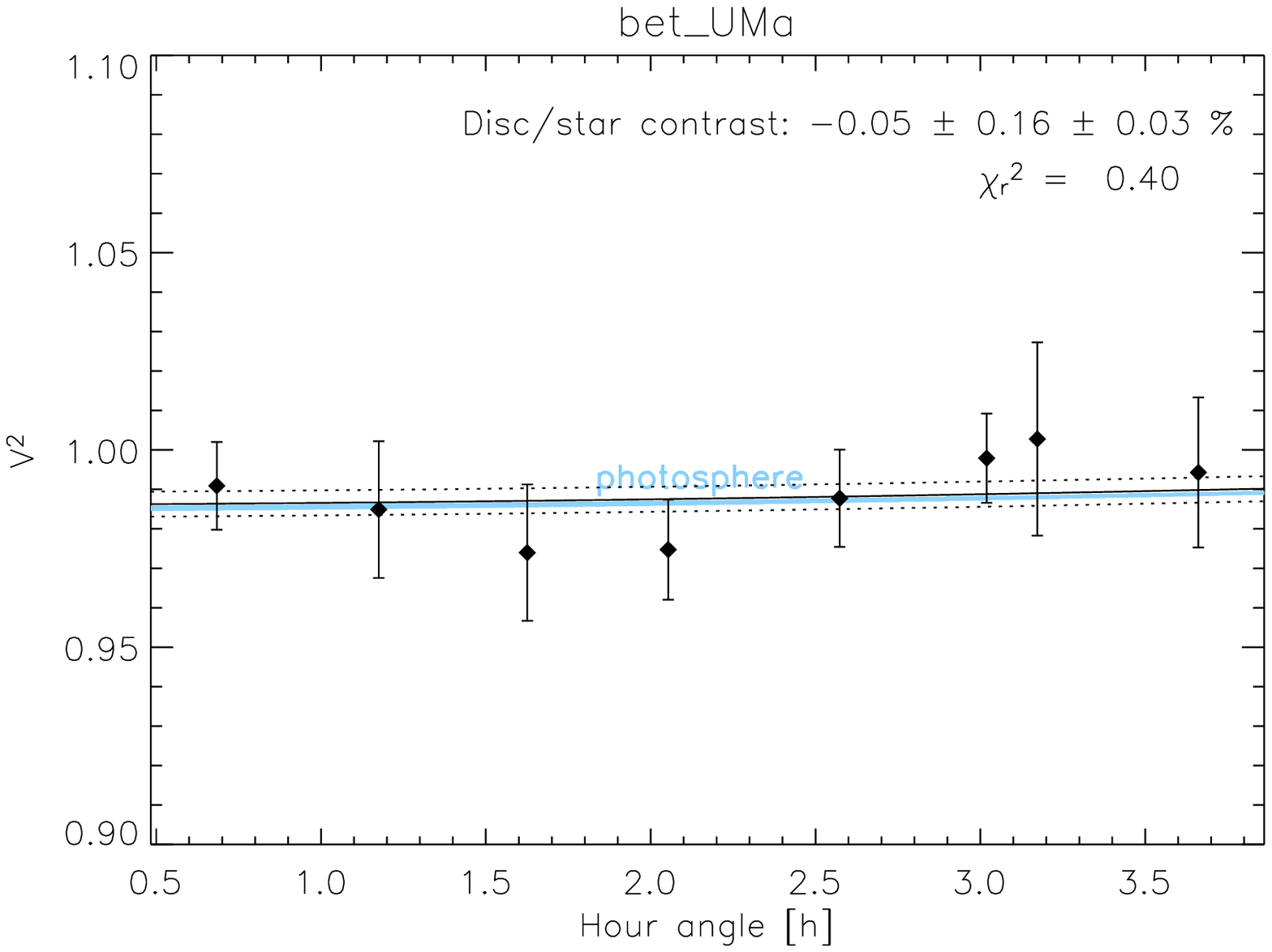} \includegraphics{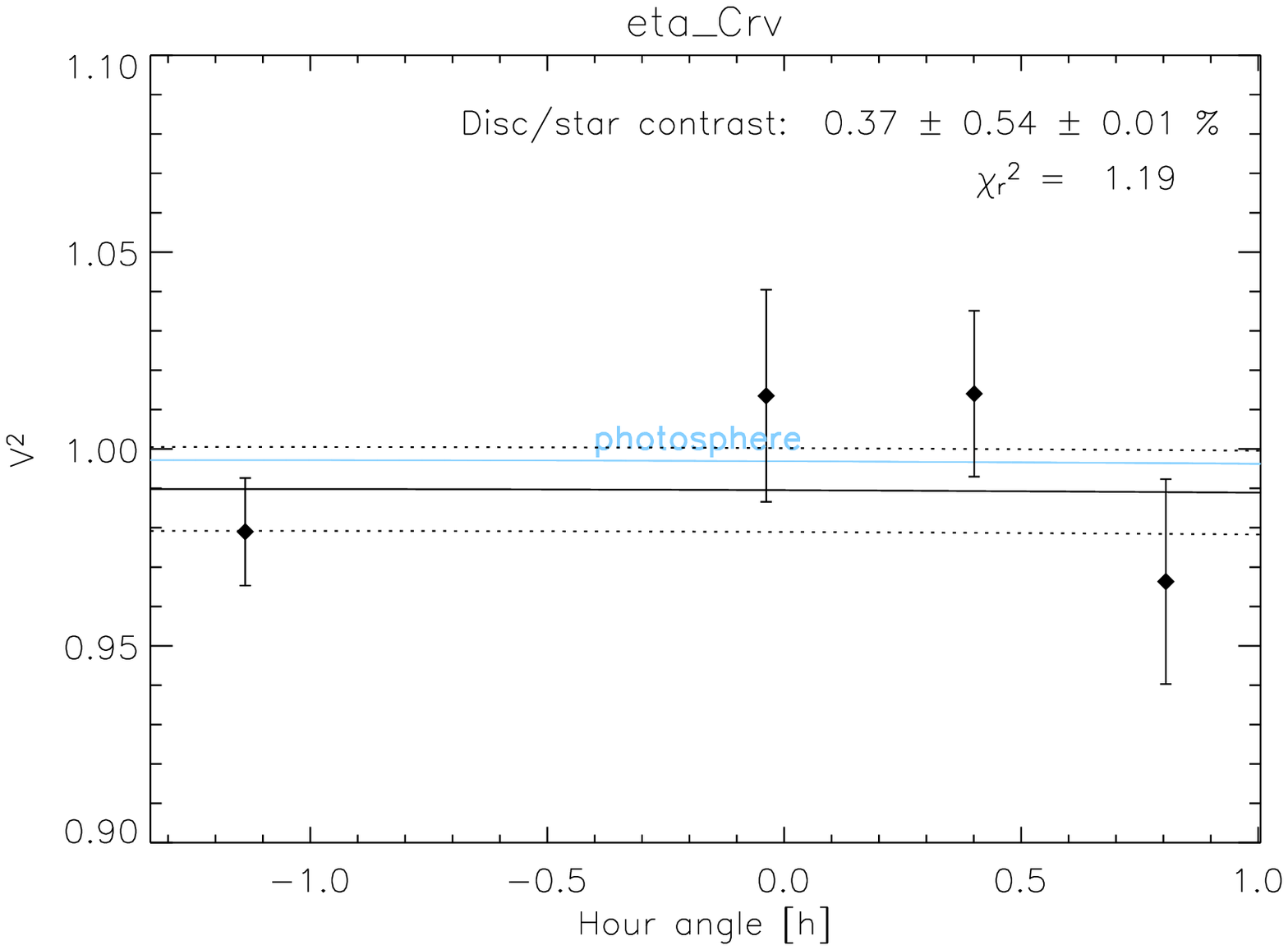}} \\
\resizebox{\hsize}{!}{\includegraphics{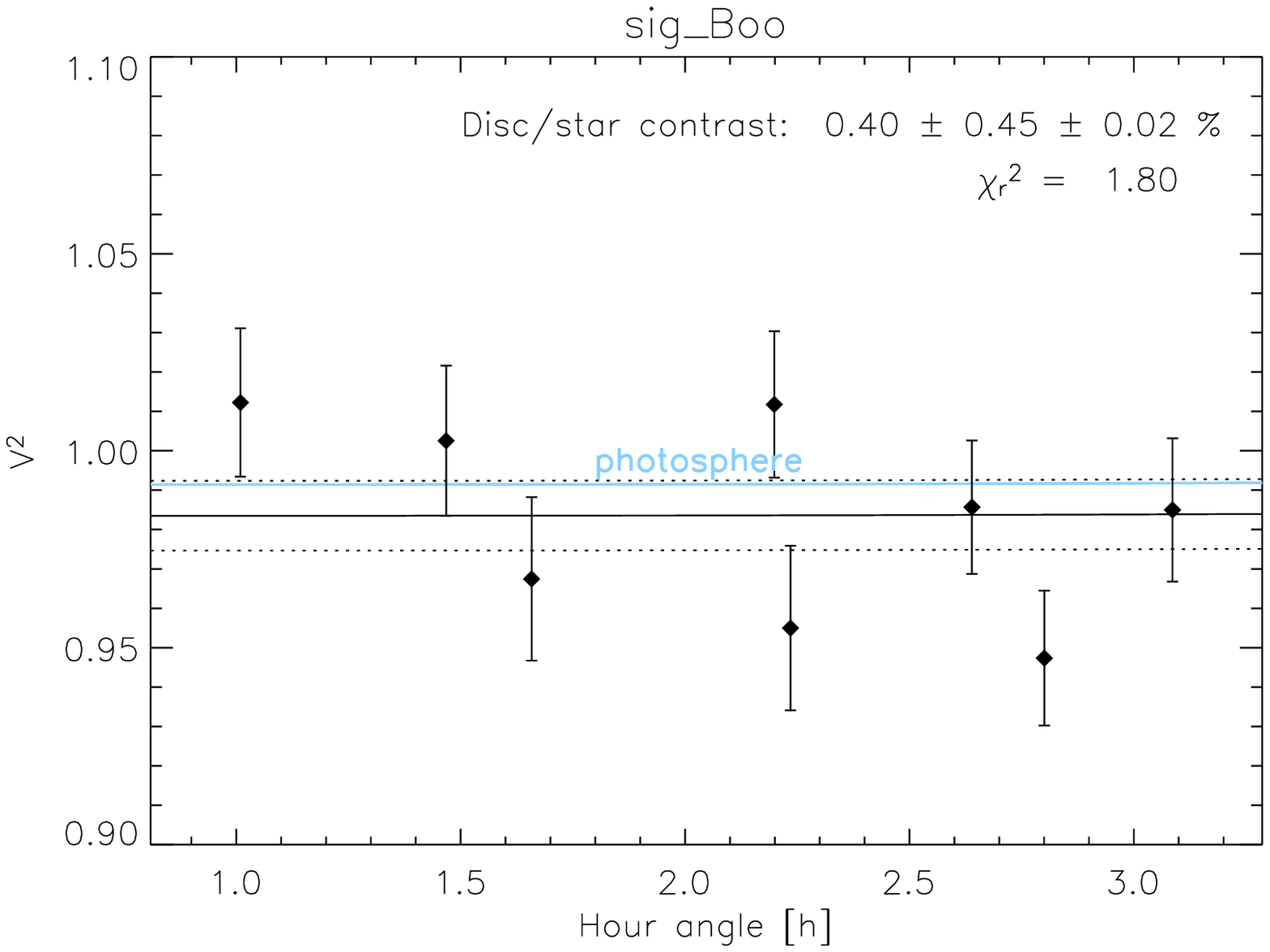} \includegraphics{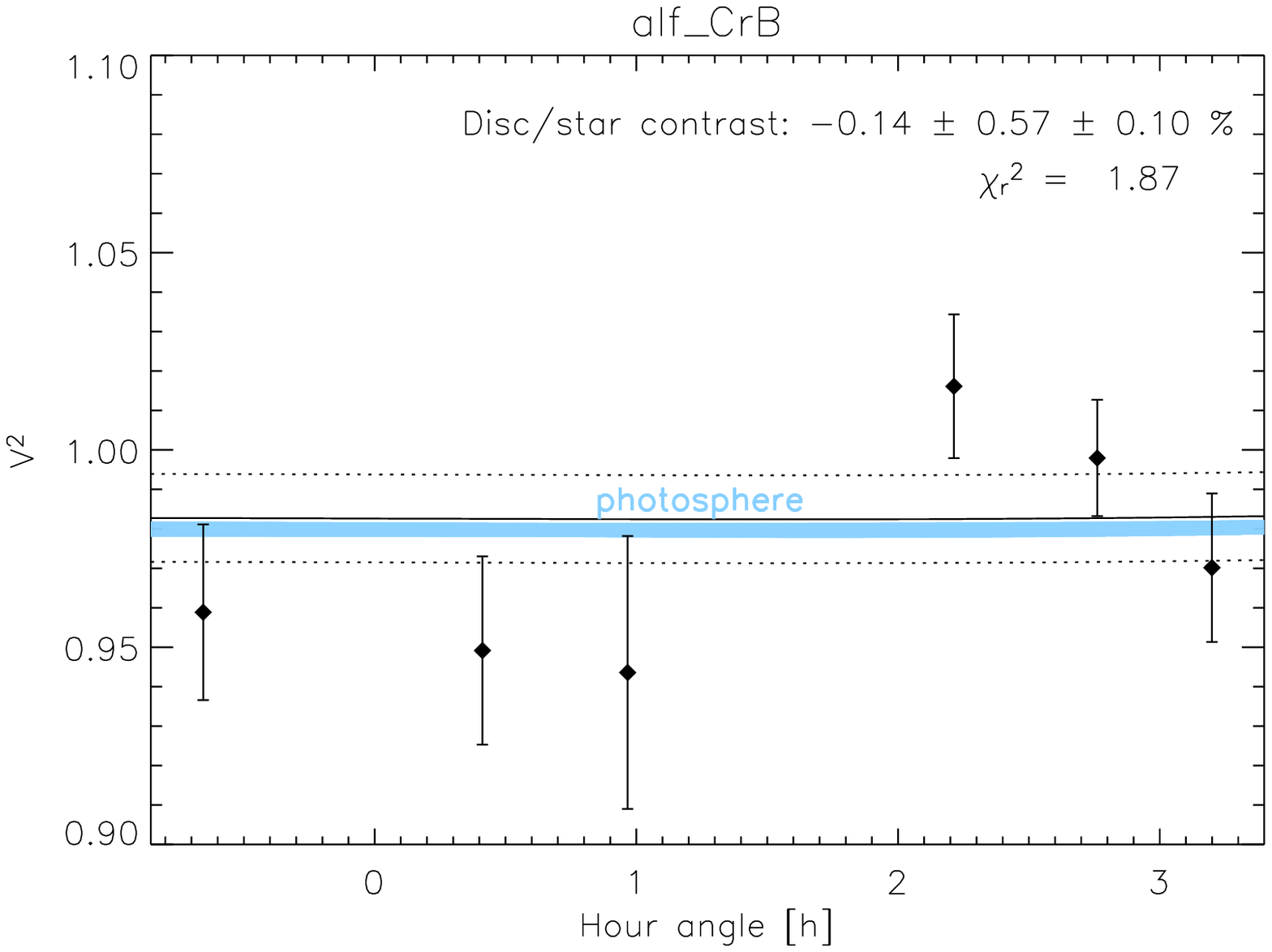}} \\
\resizebox{\hsize}{!}{\includegraphics{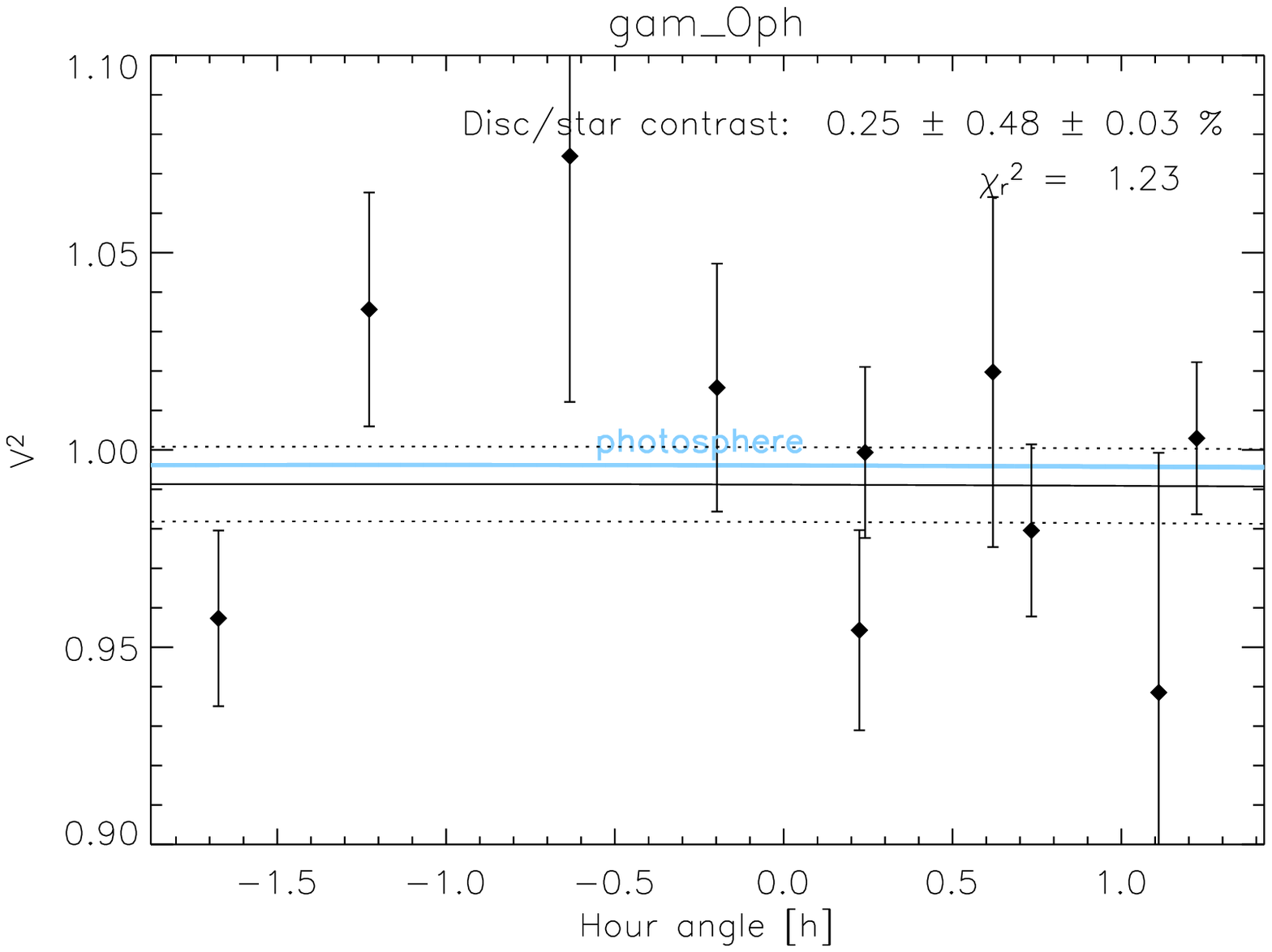} \includegraphics{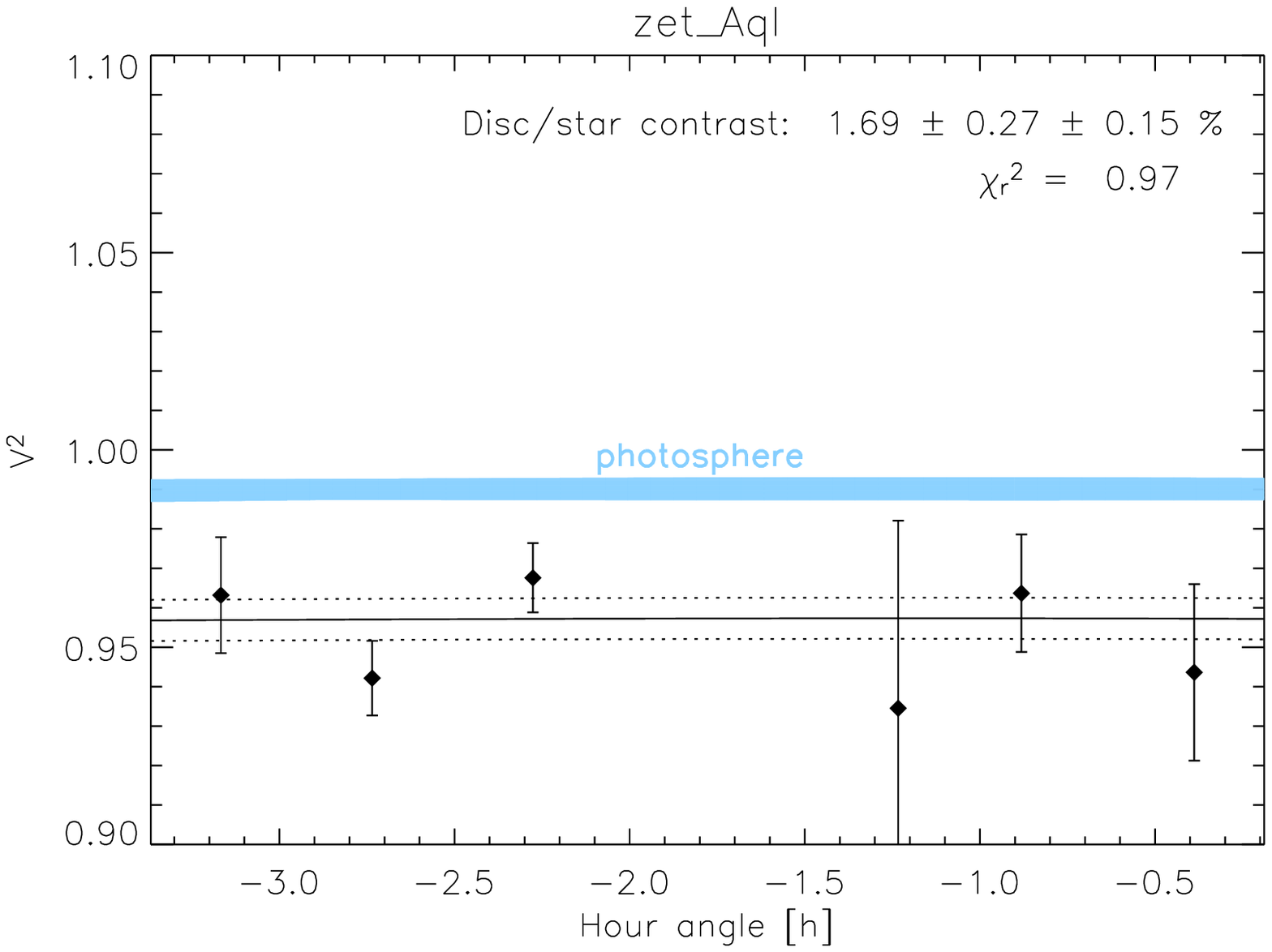}}
\caption{Squared visibilities with error bars as a function of hour angle, recorded for our six
targets on the CHARA S1--S2 baseline. The blue region represents the 1$\sigma$ box defined by our
photospheric models. We have fitted to our data a model of a limb-darkened photosphere surrounded
by a uniform circumstellar emission. The best-fit model is represented by the black solid line,
while the dotted lines represent the 1$\sigma$ statistical uncertainty on the best-fit model.
Circumstellar emission is detected when the blue region lies significantly above the uppermost
dotted line. The inferred contrast between the uniform circumstellar emission and the star is given
as inset for each star (``disc/star contrast''), together with its two error bars (the first one is
related to the statistical dispersion of the calibrated data set and the second one to the
uncertainty on the photospheric model) as well as the reduced $\chi^2$ of the fit.}
\label{fig:deficit}
\end{figure*}

The unknown orientation of the stellar rotation axis is taken into account by considering two
extreme cases: the first case (``large diameter'') assumes that the major axis of the elliptic
stellar photosphere is aligned with the mean orientation of the interferometric baseline projected
onto the sky plane for all observations, while the second case (``small diameter'') assumes the
minor axis to be aligned with the mean baseline orientation. We further add the $1\sigma$
uncertainty on the mean limb-darkened diameter (see Table~\ref{tab:diam}) to the large diameter
case, while we subtract it from the small diameter case. The first case leads to the smallest
visibilities, and the second case to the largest visibilities. In that way, we produce a
``$1\sigma$ box'' for the squared visibility of the photosphere at any projected baseline
orientation (or equivalently, any hour angle of observation on the S1--S2 baseline), which is
represented by the (thick) blue lines in Fig.~\ref{fig:deficit}. The combination of good precision
on the stellar model and a small baseline length---which leaves the photosphere almost
unresolved---results in a very small region of possible values for the squared visibilities of the
naked stars. The small curvature of the 1$\sigma$ box is due to the fact that the projected
baseline length is maximum at an hour angle of 0\,h and decreases before and after meridian
crossing.

    \subsection{Data quality and fitting procedure}

The calibrated ${\cal V}^2$ have been overlaid on the plots of the expected photospheric
visibilities in Fig.~\ref{fig:deficit}. The quality of the data, which can be estimated by the size
of the error bars and the dispersion of the data points, is not uniform for all targets, and
depends mostly on the stellar magnitude and on the seeing conditions during the observations. The
low SNR obtained on individual measurements for the faintest targets ($\eta$~Crv, $\sigma$~Boo and
$\gamma$~Oph) leads to the largest error bars (typically ranging between 2\% and 4\% for these
targets, depending on the particular conditions of the observations). The error bars for the
brightest targets typically range between 1\% and 2\%. The statistical dispersion of the data
points shows a healthy behaviour for all targets, as will be confirmed by the values of the reduced
$\chi^2$ obtained when fitting the data with a simple star+disc model. The relatively large scatter
in the $\alpha$~CrB data is due to poor seeing conditions (estimated $r_0 \simeq 4.5$\,cm) on the
night this target was observed.\footnote{The mean seeing for the other five targets of the May 2006
observing run is $r_0 \simeq 8$\,cm.}

The calibrated ${\cal V}^2$ data have been fitted with a model consisting of a limb-darkened
stellar photosphere surrounded by a uniform circumstellar emission filling the whole field-of-view
of the instrument. To account for stellar oblateness in our fit, we have defined for all baseline
orientations a mean photospheric diameter as the arithmetic mean of the angular diameters in the
``large diameter'' and ``small diameter'' cases (which both depend on the orientation of the
projected baseline). The only parameter to be fitted is then the flux ratio $\epsilon_{\rm CSE}$
between the integrated circumstellar emission and the photosphere. The results of the fitting
procedure are displayed as insets in Fig.~\ref{fig:deficit} together with the reduced $\chi^2$. The
first term in the error budget accounts for the statistical dispersion of the data, while the
second term is related to the uncertainty on the photospheric model, which includes the
uncertainties on both the surface-brightness model and the orientation of the stellar rotation
axis.

    \subsection{Results} \label{sub:results}

\begin{figure}[t]
\centering \resizebox{\hsize}{!}{\includegraphics{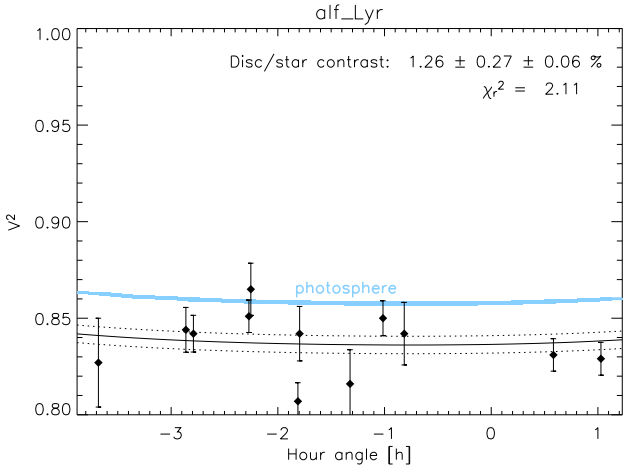}} \caption{Same as
Fig.~\ref{fig:deficit} in the case of $\alpha$~Lyr. The previously published data have been
re-reduced with the latest version of the FLUOR Data Reduction Software to produce this plot, using
the latest limb-darkened photospheric model of \citet{Aufdenberg06} as a reference.}
\label{fig:vega}
\end{figure}

Among the seven early-type stars that have been surveyed so far, two stars ($\zeta$\,Aql and
$\alpha$\,Lyr) show evidence for the presence of circumstellar emission: the fit to their
interferometric measurements falls significantly below the expected squared visibility of the
photosphere (see Fig.~\ref{fig:deficit} and Fig.~\ref{fig:vega}). The five remaining stars in our
sample show no evidence for circumstellar emission: the estimated flux ratio $\epsilon_{\rm CSE}$
between the integrated circumstellar emission and the photosphere ranges between $-0.3\sigma$ and
$+0.9\sigma$, where $\sigma$ represents the square sum of the statistical and systematic errors
bars displayed in Fig.~\ref{fig:deficit}. It must be noted that our non-detections have a
statistically-speaking healthy behaviour, with a mean $K$-band excess amounting to $0.3\sigma$ and
a standard deviation of $0.6\sigma$. This gives us good confidence in our positive detections,
which are obviously not related to a systematic effect since our stellar sample is quite
homogeneous in spectral types and magnitudes. The range of values for the reduced $\chi^2$ of the
fits is satisfactory for all cases. Let us discuss the most interesting targets individually.

\paragraph{$\zeta$\,Aql.} Our analysis of the new CHARA/FLUOR data brings to light the presence of
circumstellar emission around $\zeta$~Aql (see Fig.~\ref{fig:deficit}). The quality of the fit for
$\zeta$~Aql is quite good ($\chi^2_r=0.97$), and the final value for the $K$-band flux ratio is
$1.69\% \pm 0.31\%$, which represents a 5.5$\sigma$ detection. This result does not strongly depend
on the actual limb darkening coefficient $u_K$ of the $\zeta$~Aql photosphere: when using extreme
values of 0.0 and 0.5 for $u_K$, the final value for the $K$-band excess varies only from 1.68\% to
1.72\%. As discussed by \citet{Absil06}, this estimation of the $K$-band flux ratio does not depend
on the morphology of the circumstellar emission to the first order, as long as it has an extension
larger than the angular resolution of the interferometer (about 7\,mas in the $K$ band for a 34-m
baseline).

\paragraph{$\alpha$\,CrB.} Let us assess whether the eclipsing binary nature of the object
could significantly affect our measured visibilities. To reduce as much as possible the effect of
the companion on the measured visibilities, we have used the well-characterised ephemeris of the
system \citep{Schmitt98} to schedule the observations of $\alpha$~CrB during an eclipse. As
discussed in Sect.~\ref{sub:diam}, the expected $K$-band magnitude difference between
$\alpha$~CrB~A and its companion is $\Delta K=3.25$, which is equivalent to a contrast of about
5\%. The companion could thus potentially be at the origin of the fluctuations of about 10\%
observed in the calibrated squared visibilities. While the secondary eclipse would have been ideal
to suppress the effect of the companion, time and weather constraints have forced us to observe the
system during primary eclipse (when the companion passes in front of the primary). The $\alpha$~CrB
system was observed on April 30th from 8:10\,UT to 12:04\,UT, while the primary eclipse took place
between 6:56\,UT (start of ingress) and 21:19\,UT (end of egress). Totality of the eclipse actually
started at 9:36\,UT, so that part of the observations were obtained during ingress. Assuming that
the companion is a standard G5V star (see Sect.~\ref{sub:diam}), and using for the primary the
linear radius of $2.94\, R_{\odot}$ deduced from the surface-brightness relationships, the diameter
ratio between the two components is equal to 0.31. With the estimated $K$-band contrast, the
expected variation of squared visibility for the binary system on the S1-S2 baseline between the
start of ingress and the totality is only about 0.6\%. This effect is clearly not sufficient to
explain the dispersion of the data points, which is more likely related to the poor atmospheric
conditions during the night of April 30th. Due to our optimised observation timing, the presence of
the companion does not bias the search for a circumstellar excess at a level larger than about
0.2\% on the disc/star contrast.

\paragraph{$\eta$\,Crv.} The absence of significant $K$-band excess around this target comes rather
as a surprise, since previous mid-infrared observations have suggested the presence of warm dust
\citep{Wyatt05}, and subsequent modelling an LHB-like event \citep{Wyatt07a}. However, we note that
(i) the latest model of the $\eta$\,Crv circumstellar disc predicts a contrast larger than 1:$10^5$
in the $K$ band between the disc and the photosphere \citep{Smith08} and (ii) the small number of
data points that we have obtained on this target does not allow a high sensitivity to be reached
(only 1:62 at $3\sigma$). Therefore, our non-detection is not incompatible with previous
measurements and models, and shows that the presence of {\em warm} dust is not necessarily
accompanied by large amounts of {\em hot} dust.

\paragraph{$\alpha$\,Lyr.} The data collected in 2005 on $\alpha$\,Lyr have been re-reduced with
the latest version of the FLUOR DRS, and using a more realistic model for the stellar photosphere
based on the \citet{Aufdenberg06} model \citep[a simple uniform disc model was used in the original
study by][]{Absil06}. The new fit, displayed separately in Fig.~\ref{fig:vega}, shows a good
agreement with the previously published $K$-band excess ($1.29\% \pm 0.19\%$). Only the error bar
on the disc/star contrast has significantly changed, due to the slightly larger error bars obtained
on individual data points with the latest version of the FLUOR DRS. The contribution of the
uncertainty on the photospheric model is negligible (0.06\%) thanks to the accurate stellar model
derived from long-baselines measurements. The agreement between the result presented here and the
original disc/star contrast estimation can be seen as a welcome consistency check, since a new
version of the FLUOR DRS was used to reduce the data and a more realistic stellar model was
adopted. This also shows that the details of the photospheric model do not matter at the short
baselines considered here.


\section{Possible nature of the circumstellar emission around $\zeta$\,Aql}

The nature of the circumstellar emission detected around $\alpha$~Lyr has already been discussed by
\citet{Absil06}, showing that the presence of hot circumstellar dust is the most probable
explanation. Here, we focus on the new detection ($\zeta$~Aql), and extend our analysis to other
potential contributors to circumstellar emission, including stellar winds and mass loss events. The
following discussion is based on our interferometric observations, complemented only by archival
infrared photometric measurements. In Sect.~\ref{sec:discussion}, we will include additional data
from other instruments to further constrain the nature of the circumstellar emission. In this way,
we want to make clear what type of information can actually be derived from a limited amount of
squared visibilities obtained on a short two-telescope baseline.

The only strong constraints that we derive on the circumstellar emission from our interferometric
measurements is that it represents 1.69\% of the stellar flux in $K$ band\footnote{The effective
wavelength of the FLUOR observation is 2.132\,$\mu$m.} and that it is located within the FLUOR
field-of-view, which has a FWHM of $0\farcs8$. Due to the quasi-Gaussian beam of the FLUOR
single-mode fibres, circumstellar emission located at the edge of the field (e.g., dust ring or
point-like source) needs to be significantly brighter than 1.69\% to produce the detected
visibility deficit. Taking the transmission at the centre of the field as a reference, only half of
the flux emitted at $0\farcs4$ (i.e., at a projected distance of 9\,AU) makes it through the
fibre's transmission pattern. At 30\,AU, the flux transmission decreases to 10\%, following the
theoretical Gaussian profile.

    \subsection{The debris disc scenario} \label{sub:disc}

The first scenario that we wish to investigate is the presence of hot circumstellar dust within the
field-of-view of the FLUOR instrument, which would be at the origin of the measured visibility
deficit. Although the origin and survival of hot dust grains close to an early-type MS star is a
matter of debate, a plausible debris disc model, reproducing the interferometric measurements as
well as archival photometric measurements in the near- and mid-infrared has been proposed for
$\alpha$~Lyr \citep{Absil06}. In that paper, we also showed that the actual morphology of the
circumstellar dust disc does not have a significant influence on the visibility deficit measured by
CHARA/FLUOR: only the flux ratio between the integrated disc emission and the stellar photosphere
actually matters.

In this section, we investigate whether a debris disc model similar to that proposed for
$\alpha$\,Lyr would be suitable to reproduce the available measurements of $\zeta$~Aql. The purpose
is not to propose a ``best-fit'' set of physical parameters for the dust disc, but only to check
whether there exists at least a debris disc model consistent with the near- and mid-infrared
observations of $\zeta$~Aql. The question of the origin and physical properties of such dust grains
will be discussed in a forthcoming paper.

To perform our simulations, we have complemented our direct measurement of the $K$-band excess flux
with archival spectro-photometric measurements at near- and mid-infrared wavelengths (see
Table~\ref{tab:photo}). In addition to the measurements found in the literature, we have used an
IRS spectrum found in the Spitzer archives, which we have reduced with the c2d pipeline developed
by \citet{Lahuis06}. The spectrum has been binned into a few equivalent broad-band photometric
measurements for the sake of SED modelling (see Table~\ref{tab:photo}). Because all these fluxes
are valid for the whole star/disc system, they have been converted into an estimated excess flux by
subtracting the photospheric flux. The photospheric emission model is chosen from the NextGen grid
\citep{Hauschildt99}, using stellar parameters as close as possible to the parameters estimated by
\citet{Gray03} and \citet{Adelman02}, i.e., $T_{\rm eff}=9200$\,K, $\log g = 4.00$ and ${\rm
M}/{\rm H}=0.0$. The NextGen model is scaled to match the observed $V=2.988$, which results in an
estimated luminosity of $39 L_{\odot}$, in agreement with the estimation of \citet{Malagnini90}. We
have estimated the error on the NextGen photospheric infrared fluxes by comparing neighbouring
models on the NextGen grid corresponding to the typical range in $T_{\rm eff}$ and $\log g$
estimations found in the literature. Assuming effective temperatures comprised between 9200\,K and
9400\,K, and surface gravities between 3.5 and 4.0, we have derived a typical error of 2\% on the
estimated fluxes between 1\,$\mu$m and 32\,$\mu$m.

\begin{table}[t]
\caption{Available constraints on the near- and mid-infrared photometry of $\zeta$\,Aql.
References: (1) \citet{Morel78}; (2) \citet{Leggett86}; (3) \citet{Egan03}; (4) \citet{Chen05}. The
archival Spitzer-IRS spectrum has been binned into a few photometric data points. Photometric data
have been compared to our NextGen photospheric model of $\zeta$\,Aql to derive fractional excesses,
taking into account the typical uncertainty of 2\% on the photospheric flux in the infrared (this
uncertainty has been added to the estimated errors on the measured fluxes to compute the errors on
the excesses).} \label{tab:photo} \centering
\begin{tabular}{cccl}
\hline \hline  Wavelength   & $F_{\rm meas}$ (Jy)& Excess            & Ref.
\\ \hline      1.2\,$\mu$m  & $112.4 \pm 4.1$    & $-4.5 \pm 5.5$\%  & (1)
\\             1.24\,$\mu$m & $114.5 \pm 2.1$    & $2.0 \pm 3.9$\%   & (2)
\\             2.13\,$\mu$m &       ---          & $1.69 \pm 0.31$\% & CHARA/FLUOR
\\             4.35\,$\mu$m & $13.3 \pm 1.7$     & $-2.3 \pm 14.7$\%   & (3)
\\             5.5\,$\mu$m  & $9.18 \pm 0.42$    & $5.3 \pm 6.8$\%   & IRS spectrum
\\             7.5\,$\mu$m  & $4.85 \pm 0.22$    & $1.7 \pm 6.6$\%   & IRS spectrum
\\             8.28\,$\mu$m & $4.34 \pm 0.18$    & $9.5 \pm 6.5$\%   & (3)
\\             12.13\,$\mu$m& $2.15 \pm 0.13$    & $14.8 \pm 8.9$\%   & (3)
\\             12.5\,$\mu$m & $1.73 \pm 0.10$    & $-1.8 \pm 7.6$\%   & IRS spectrum
\\             14.65\,$\mu$m& $1.44 \pm 0.10$    & $12.1 \pm 9.8$\%   & (3)
\\             18\,$\mu$m   & $0.868 \pm 0.051$  & $2.0 \pm 8.0$\%   & IRS spectrum
\\             24\,$\mu$m   & $0.475 \pm 0.048$  & $-0.7 \pm 12.0$\%   & (4)
\\             32\,$\mu$m   & $0.270 \pm 0.025$  & $0.8 \pm 11.5$\%   & IRS spectrum
\\ \hline
\end{tabular}
\end{table}

\begin{figure*}[t]
\centering \resizebox{\hsize}{!}{\includegraphics[angle=90]{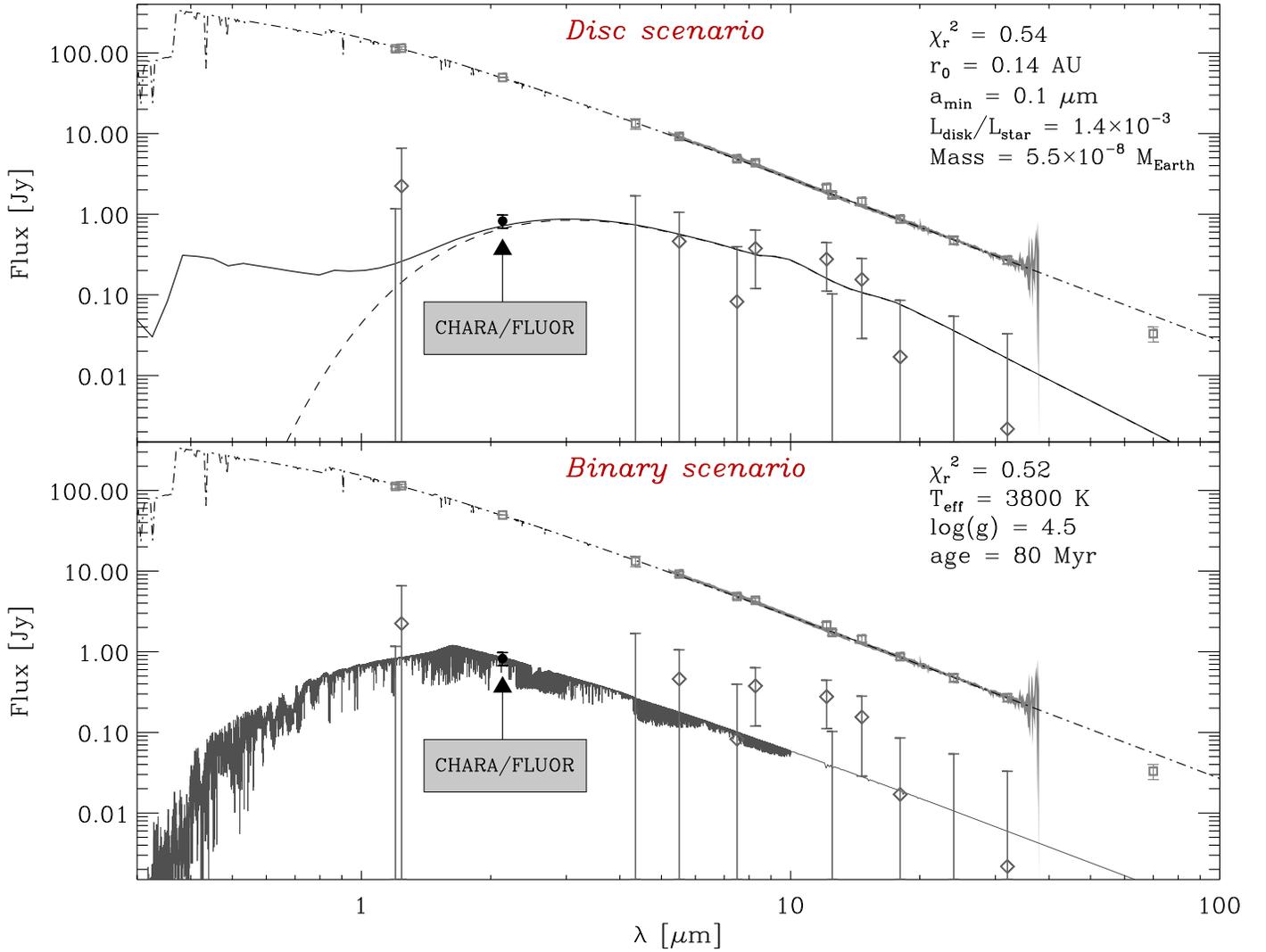}} \caption{Top panel: A
possible fit of our debris disk model to the photometric and interferometric constraints of
Table~\ref{tab:photo}. The dashed line represents the contribution of thermal emission to the
overall disc emission, while the solid line includes the scattered light contribution. Photometric
measurements are represented by squares with error bars, and the related excesses by diamonds. The
CHARA/FLUOR interferometric measurement is represented by a filled circle with error bar. The whole
IRS spectrum is represented by a grey box with a width equal to its error bar, and was converted
into a few regularly spaced (in log scale) broad-band photometric measurements for the fitting
procedure. The MIPS photometric measurement at 70\,$\mu$m \citep{Chen05} is also represented but
was not included in the fit. Bottom panel: Same figure in the case of a low-mass companion. The
spectrum of the companion is represented by a NextGen model \citep{Hauschildt99} where the stellar
parameters ($T_{\rm eff}=3800$\,K, $\log g = 4.5$) have been chosen as close as possible to our
best-fit model (see Sect.~\ref{sub:binary}). The resolution of the NextGen spectrum of the A0-type
host star has been degraded by a factor 20 in both panels for the sake of clarity.} \label{fig:sed}
\end{figure*}

Our debris disc model is an adaptation of the ``best-fit'' model that we derived in the case of
$\alpha$\,Lyr \citep{Absil06}, based on the model of \citet{Augereau99}. It assumes a size
distribution $dn(a) \propto a^{-3.7} da$ with limiting grain sizes $a_{\rm min} = 0.1$~$\mu$m and
$a_{\rm max} = 1500$~$\mu$m, a surface density power-law $\Sigma(r) \propto r^{-4}$, and chemical
composition of 50\% amorphous carbon and 50\% glassy olivine. The inner radius of the dust
distribution has been recomputed in a similar way as in the case of $\alpha$\,Lyr, and is found to
be equal to 0.14\,AU. This corresponds to the sublimation radius for the grains larger than about
1\,$\mu$m, while 0.1\,$\mu$m grains sublimate at $\sim 0.45$\,AU in the model. All grains within
their sublimation distance are self-consistently eliminated from the simulations, leading to
regions around the inner disc edge depleted in submicronic grains. The only free parameter is the
total mass of the disc, which has been adjusted to the photometric and interferometric data in
Fig.\ref{fig:sed}. With a reduced $\chi^2$ of 0.54, our model does a good job in reproducing the
global SED of $\zeta$\,Aql from 1 to 32\,$\mu$m, and demonstrates that the presence of hot dust,
with a total dust mass of only $5.5\times10^{-8}$\,M$_{\oplus}$ and a fractional luminosity of
about $10^{-3}$, is a viable explanation to the observed $K$-band excess. It must be noted that, as
in the case of $\alpha$\,Lyr, most of the grains in this model are submicronic and located close to
their sublimation radius. Possible scenarios to explain the presence of such grains have already
been proposed \citep{Absil06} and will be further discussed in a forthcoming paper.

The hot debris disc model proposed for $\zeta$\,Aql can further be used to derive upper limits on
the mass of hot dust around the stars in our sample for which no detection was reported. This
estimation is based on the fact that none of the surveyed stars shows a significant photometric
infrared excess in the 1--5\,$\mu$m region, and on the fact that their spectral types are close
enough for the same debris disc model to apply to all of them. Noting that the final error bar on
the disc/star flux ratio is typically 0.5\% for these stars (except for $\beta$\,UMa, which has a
significantly lower error bar of 0.16\%), we derive a typical $3\sigma$ upper limit of
$5\times10^{-8}$\,M$_{\oplus}$ for the hot dust content within the FLUOR field-of-view of
$0\farcs8$ (i.e., 18\,AU at the mean distance of our sample). Scaling this mass to the smaller
error bar of $\beta$\,UMa gives a $3\sigma$ upper limit of $1.5\times10^{-8}$\,M$_{\oplus}$ for
this particular system. We note that our observations are only sensitive to hot dust, and that
larger amounts of colder dust could be present, especially beyond a few AUs where equilibrium
temperatures are significantly smaller than 1000\,K (see e.g.\ the discussion of $\eta$\,Crv in
Sect.~\ref{sub:results}).

    \subsection{Stellar wind and mass loss} \label{sub:winds}

With their relatively high masses and high rotational velocities, one may think that our target
stars could be the subject of significant mass loss or winds that could, as in the Be phenomenon,
lead to a substantial near-infrared emission in addition to the photospheric level. However, winds
of A-type stars are actually suspected to be very weak, as the mechanisms which are efficient in
both later type stars and earlier ones are not expected to operate on A-type stars: radiative
pressure is very small (compared to B-type stars), and there is no deep convection giving rise to
activity phenomenon as in late-type stars \citep{Babel95}. Convection is actually expected to start
for effective temperatures below 6500~K \citep{Lamers99}, i.e., for spectral types later than about
F6\,V, so that A-type stars are not expected to have chromospheres and coronae. This fact seems to
be corroborated by the lack of observational evidence for winds emanating from A-type stars
\citep{Lanz92,Landstreet98}, as well as by the absence of X-ray emission from most A-type stars
\citep{Schroder07}.

Nonetheless, we note that the presence of a stellar wind has been detected around Sirius (A1\,V)
through the observation of an absorption feature in the blue wings of the Mg\,{\footnotesize II}
and H\,{\footnotesize I} resonance lines \citep{Bertin95}. The inferred mass-loss rate ranges
between $2\tothe{-13}$ and $1.5\tothe{-12} M_{\odot} {\rm yr}^{-1}$, and has not been shown to
produce any significant near-infrared emission. Based on radiatively driven wind models,
\citet{Babel95} predicts much lower mass loss rates ($<10^{-16} M_{\odot} {\rm yr}^{-1}$), with
winds consisting of only metals. Additionally, the infrared excess that could be associated to the
winds from hot stars is expected to originate mostly from a region $R<1.5R_{\ast}$
\citep{Lamers99}, which does not match our observations because such a compact emission would not
produce a significant visibility reduction at the short baselines used here.

Intriguingly, shell lines of Ti\,{\scriptsize II} have been detected by \citet{Abt97} in the
ultraviolet spectra of approximately one-quarter of the most rapidly rotating normal A-type dwarfs
($v \sin i \ge 200$~km\,s$^{-1}$). In that paper, the authors suggest that this ``hot disc''
phenomenon might be related to sporadic mass-loss events and does not seem to correlate to the
presence of cold dust in an outer debris disc. However, more recently, \citet{Abt04} and
\citet{Abt08} have argued that these hot discs are actually accreted from the interstellar medium.
The argument is based on the fact that such discs rarely occur around stars within the heart of the
Local Interstellar Bubble (LIB). The author proposes a model in which stars accrete discs in dense
interstellar regions, but loose them in regions of low interstellar density, such as the LIB, where
the stellar winds exceed the accretion rate. The spectral lines associated with the hot disc
disappear on timescales of roughly decades. Since our target stars are all located well within the
LIB ($\ll 100$\,pc), this phenomenon does not seem appropriate to reproduce our observations.

All these elements indicate that a stellar mass-loss origin to the observed near-infrared excess is
highly unlikely in the case of A-type dwarfs. Furthermore, we have examined archival HARPS data on
$\zeta$\,Aql and found no sign of H$\alpha$ emission, which accompanies strong stellar winds
produced by hot stars. The stellar wind scenario will thus be discarded in the rest of the
discussion.




    \subsection{Point-like source} \label{sub:binary}

Because of our sparse sampling of spatial frequencies, we cannot determine the actual morphology of
the excess emission source, and a point-like source located within the FLUOR field-of-view (either
a bound companion or a background source) could be at the origin of the observed visibility
deficit. To assess the effect of a faint companion on the measured visibilities, two cases must be
considered, depending whether the two fringe packets are superimposed or not. Taking into account
the coherence length of the FLUOR fringes (about 25~$\mu$m), the minimum on-sky angular separation
for two separated fringe packets is 150\,mas along the baseline direction, i.e., about 3.5~AU at
the distance of $\zeta$~Aql. For larger separations, the off-axis point-like source contributes
essentially as an incoherent emission and produces a squared visibility deficit $\Delta{\cal V}^2 =
2f$, where $f$ is the flux ratio between the primary and the secondary. A point-like source within
the FLUOR field-of-view, located at a distance larger than 150\,mas from $\zeta$\,Aql in the
projected direction of the interferometric baseline and producing 1.69\% of the photospheric flux
of $\zeta$\,Aql, would therefore be a possible explanation to the measured visibility deficit.

Let us now investigate the case of a close companion ($<150$\,mas), for which the two fringe
packets would be superposed. In that case, the companion produces a modulation in the fringe
visibility depending on its position and on the geometry of the array, which changes as the Earth
rotates. The squared visibilities are therefore expected to be modulated as a function of the hour
angle of the observation. To determine which combination of angular separation and position angle
would be compatible with the interferometric observations, we compute the squared visibilities for
a binary system at various distances and position angles, assuming that the off-axis source did not
move during the 11 nights of our May 2006 observing run.\footnote{This hypothesis is valid for
background sources and for bound companions with semi-major axes larger than 0.7\,AU (which does
not prevent the companion from being at a projected angular separation as small as a few mas from
the target star).} For each given position, the contrast between the two components is then
adjusted to minimise the $\chi^2$ distance between the synthetic and measured squared visibilities.
In this way, we produce a $\chi^2$ map, taking into account the wide bandwidth effect of the FLUOR
instrument.

\begin{figure}[t]
\centering \resizebox{\hsize}{!}{\includegraphics{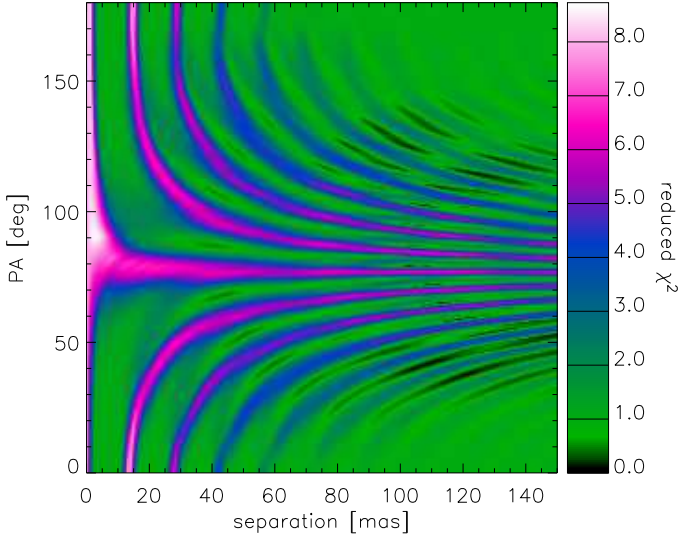}} \caption{Map of
the reduced $\chi^2$ distance between a binary star model and our CHARA/FLUOR squared visibilities
for $\zeta$\,Aql, as a function of the position of the faint companion. The binary flux ratio has
been adjusted separately at each given position to minimise the $\chi^2$. Dark regions indicates
binary solutions compatible with the CHARA/FLUOR data.} \label{fig:chi2}
\end{figure}

The $\chi^2$ map, represented in polar coordinates in Fig.~\ref{fig:chi2}, indicates that many
different locations of the off-axis point-like source may reproduce the squared visibilities
measured by CHARA/FLUOR. In particular, for any given angular separation larger than 3\,mas, we can
always find a position angle that makes the simulated data compatible with the observed
visibilities (reduced $\chi^2 \lesssim 1$). We also note that the optimum binary flux ratio at the
positions where the reduced $\chi^2$ is smaller than 2 is in excellent agreement with the
``disc/star'' contrast computed in Sect.~\ref{sub:disc}: the mean flux ratio at such locations is
1.65\%, with a standard deviation of only 0.22\%. This further confirms that the morphology of the
circumstellar emission does not have a significant influence on the inferred flux ratio, and
indicates that the simplified computation of the binary flux ratio assuming separated fringe
packets can also be used as a good approximation for superposed fringe packets.

Figure~\ref{fig:chi2} is also a nice illustration of the wide bandwidth effect in optical
interferometry: for position angles parallel to the mean baseline orientation (i.e., ${\rm
PA}=170^{\circ}$), one can see that the ripples in the $\chi^2$ map due to the visibility
modulation produced by the binary star progressively wash out as the companion moves away from the
primary star. This effect can be interpreted as the progressive separation of the two fringe
packets. The reduced $\chi^2$ values obtained at a position where the fringe packets are
essentially separated (i.e., separation of 150\,mas and position angle close to $170^{\circ}$) make
a nice connection with the case of large separations described earlier in this section. One can
also note that for position angles perpendicular to the mean baseline azimuth, the reduced $\chi^2$
does not change significantly as a function of distance, because only the projection of the angular
separation along the baseline direction actually matters.

Finally, we have checked that we can find on the NextGen stellar atmosphere grid a low-mass stellar
companion compatible in terms of flux both with our CHARA/FLUOR measurement and with the archival
photometric data listed in Table~\ref{tab:photo}. In the bottom panel of Fig.~\ref{fig:sed}, we
show that an M0V-type star with $T_{\rm eff}=3800$\,K and $\log g=4.5$ nicely fits the whole data
set. A low-mass companion located at an angular distance larger than 3\,mas from the central star
is therefore a plausible explanation to the observed visibility deficit at short baselines.


\section{Discussion} \label{sec:discussion}

Our interferometric observations are not sufficient to determine the morphology of the
circumstellar emission. In this section, we investigate whether complementary data obtained with
various types of instruments could further constrain the nature of the circumstellar excess.

    \subsection{Further constraints on the debris disc scenario}

The identification of $\zeta$\,Aql as a debris disc star by \citet{Chen06} was based on the
measurement of significant excess emission on top of the expected photospheric flux at 12 and
24\,$\mu$m. However, Fig.~\ref{fig:sed} strongly contradicts this statement, and shows the absence
of cold circumstellar dust at the sensitivity limit of the Spitzer IRS and MIPS instruments. The
discrepancy with the analysis of \citet{Chen06} is related to the effective temperature of 11912\,K
that they have used to compute the photospheric emission of $\zeta$\,Aql, which does not match the
estimated temperatures found in the literature, ranging between 9190 and 9680\,K \citep[see
e.g.][]{Gray03,Adelman02,Malagnini90}. Only the polar temperature could reach such a high value
\citep{Peterson06a}, while $\zeta$\,Aql is view essentially equator-on. The very good match between
our photospheric emission model and the IRS spectrum gives us a high confidence in our flux
estimation, and clearly shows that no excess emission is detected in the 24--70\,$\mu$m region.
This also explains why a black body disc model did not allow \citet{Chen05} to consistently
reproduce the 24 and 70\,$\mu$m excesses that they had derived. The fact that the MIPS 70\,$\mu$m
measurement falls about $3\sigma$ below our photospheric model (see Fig.~\ref{fig:sed}) could be
either due to a poor flux calibration on this target or to an underestimated error bar.


Despite the absence of observable amounts of cold dust, the presence of hot circumstellar dust in
the close neighbourhood of 83\,Myr old A-type star could still be explained in the context of
terrestrial-planet formation models. In particular, \citet{Kenyon06} describe oligarchic and
chaotic growths in the terrestrial region around a solar-type star, and show that the planet
formation timescale is 10--100\,Myr \citep[see also][and references therein]{Chambers04}. The onset
of chaotic growth depends on the surface density, and typically ranges between 0.1\,Myr and a few
Myr. During the subsequent chaotic phase, the dynamical perturbations between oligarchs may lead to
enhanced dust production. This potential origin for the dust observed around $\zeta$\,Aql is also
supported by the study of \citet{Rieke05}, who suggest that the dust detected towards relatively
young A-type stars may be related to the planet accretion end game. According to
\citet{Chambers01}, the accretion end game is predicted to be dominated by a small number of
collisions between large bodies (e.g., the early Earth and the impactor that caused formation of
the Earth's Moon) and may stretch over 100--200\,Myr. \citet{Kokubo06} also show that the giant
impact stage (i.e., the accretion timescale) around a Solar-type star lasts for about 100\,Myr.
However, it must be noted that, according to \citet{Kenyon04}, the major part of the warm infrared
excess may disappear within 1--10\,Myr during terrestrial planet formation around Sun-like star.
Furthermore, \citet{Kenyon05} show that, in the case of a 3--20\,AU region around A-type stars,
1000\,km objects are formed within 10\,Myr, the dust luminosity peaks at 1\,Myr and is already at a
very low level at 100\,Myr.

In conclusion, even though the presence of hot dust in the close environment of $\zeta$\,Aql cannot
be ruled out, the likeliness of this scenario can be questioned due to the absence of a large
reservoir of cold dust in the outer planetary system.


    \subsection{Further constraints on the binary scenario}

\begin{figure}[t]
\centering \resizebox{\hsize}{!}{\includegraphics{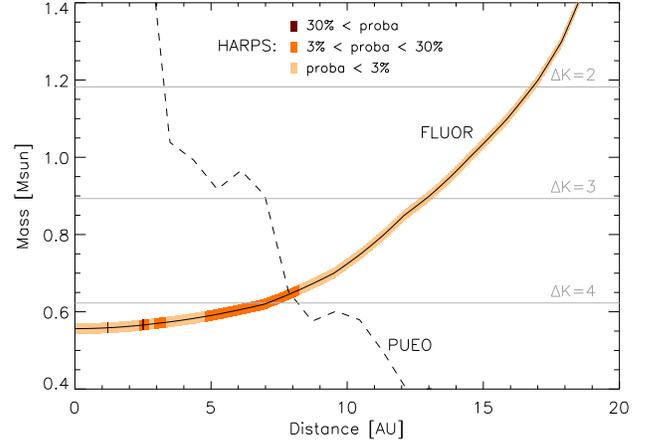}} \caption{Constraints on the mass
and distance of a hypothetic companion around $\zeta$~Aql. The solid line, labelled FLUOR,
represents the solutions that are compatible with our interferometric observations. The adaptive
optics images that we have obtained with PUEO constrain the companion to be located below the
dashed line (PUEO $3\sigma$ detection limit). The probability that the companion defined by the
FLUOR curve fits our HARPS data is represented by three colour levels along the FLUOR curve. The
correspondance between the mass of the companion and its $K$-band contrast with respect to the
primary is indicated by grey horizontal lines, based on the evolutionary models of
\citet{Baraffe98}.} \label{fig:companion}
\end{figure}

To get a better grasp at the possible nature of the potential point-like source around
$\zeta$\,Aql, we compute the flux of the hypothetical companion reproducing the observed visibility
deficit as a function of its distance, taking into account the Gaussian off-axis transmission of
the single-mode fibre. For the sake of clarity, the apparent binary flux ratio seen through the
Gaussian transmission pattern is approximated by the simple expression ($f=\Delta{\cal V}^2/2$),
which strictly applies only to wide binaries ($>150$\,mas), even though we have shown in
Sect.~\ref{sub:binary} that this relationship is a good approximation to the best-fit solutions for
close binaries. The derived flux ratio is then converted into the mass of the hidden companion in
Fig.~\ref{fig:companion} (solid line) as a function of linear distance to the central star, using
the mass-luminosity relationships of \citet{Baraffe98}.

As already discussed by \citet{Absil06} and \citetalias{DiFolco07}, the presence of a background
source producing such a flux within the $0\farcs8$ FLUOR field-of-view is highly unlikely.
Therefore, we investigate here the possible presence of a bound companion around $\zeta$~Aql.
Although $\zeta$~Aql is known to be part of a wide multiple system (ADS\,12026), its detected
companions are no closer than $5\arcsec$ so that they cannot affect our visibility measurements.
This star is classified as a spectroscopic binary in the WDS catalogue. However, this
classification is most probably wrong as it is based on a measurement from the late 19th century,
which has not been confirmed since then. The detection of X-ray emission towards $\zeta$\,Aql with
ROSAT is a possible sign for the presence of a low-mass companion \citep{Schroder07}. However, due
to the large PSF of the ROSAT satellite ($18\arcsec$), the visual companion that was located at
$6\farcs5$ from $\zeta$\,Aql in 1934 according to the WDS catalogue could be at the origin of the
observed X-rays. To our knowledge, no other sign of the presence of a close companion to
$\zeta$~Aql has been reported in the literature. However, close and faint companions are difficult
to detect.

In order to improve our knowledge of the close environment of $\zeta$~Aql, deep adaptive optics
images have been obtained in the $K$ band with the PUEO instrument at the Canada-France-Hawaii
Telescope. Under good seeing conditions (about $0\farcs6$), no companion was detected at the
sensitivity limit of the instrument ($\Delta K=4$ at $0\farcs3$, and improving for larger
separations). The $3\sigma$ sensitivity to point-like companions has been computed as a function of
angular separation and has been converted into the maximum mass of a hidden companion in
Fig.~\ref{fig:companion} (dashed line) as a function of linear distance, using the mass-luminosity
relationships of \citet{Baraffe98}. All solutions with companions orbiting further than about 8\,AU
can be discarded by the PUEO observations, but closer companions are not constrained.

\begin{figure}[t]
\centering \resizebox{\hsize}{!}{\includegraphics{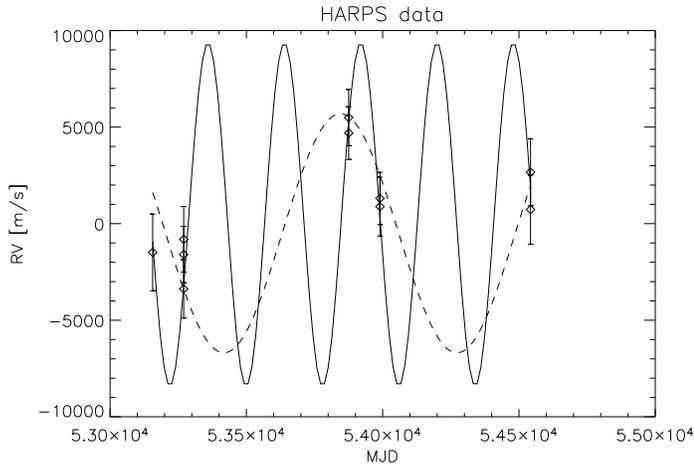}} \caption{Radial velocities
measured on $\zeta$\,Aql with HARPS during the past 4 years. The two most probable circular orbits
compatible with both the HARPS data and the FLUOR constraints have been overlaid as solid and
dotted curves.} \label{fig:harps}
\end{figure}

We have therefore used radial velocity data obtained with HARPS to further constrain the possible
presence of a low-mass companion around $\zeta$\,Aql. We have tried to fit simple orbital solutions
to the HARPS data collected between 2004 and 2008 (represented in Fig.~\ref{fig:harps}), assuming a
90$^{\circ}$ inclination for the binary system and a zero eccentricity. Each point along the FLUOR
curve in Fig.~\ref{fig:companion} defines a single couple ``period -- maximum radial velocity'',
which is computed with classical orbital dynamics relationships and defines the shape of the
sinusoidal radial velocity curve. For each couple ``period -- velocity'' on the FLUOR curve, we
have computed the $\chi^2$ between the observed HARPS data and a series of sinusoidal models with
various time and radial velocity offsets. The lowest $\chi^2$, associated to the best-fit offsets
in time and radial velocities, has then been converted into a probability, which represents the
likeliness that the binary star model actually reproduces the HARPS data. Probabilities are
represented by three colour levels in Fig.~\ref{fig:companion}, and show that two solutions at
orbital radii $a=1.2$\,AU and $a=2.5$\,AU are most probable (i.e., 47\,mas and 98\,mas
respectively). The potential companions have respective masses of 0.56 and $0.57\,M_{\odot}$ and
complete their orbits in 280 and 842 days. These two most probable solutions are respectively
represented by solid and dotted curves in Fig.~\ref{fig:harps}. The closest model on the
\citet{Baraffe98} evolutionary grid at 80\,Myr for these two potential companions is a
$0.57\,M_{\odot}$ object with an effective temperature $T_{\rm eff}=3779$\,K and a surface gravity
$\log g =4.68$. Its absolute magnitude is $M_K=5.22$, giving $K=7.26$ at the distance of
$\zeta$\,Aql, and thus a $\Delta K = 4.36$ with the primary. The associated $V$-band magnitude is
11.2, so that the contrast with the primary is $\Delta V=8.2$.

Even though the HARPS data do not give a definitive answer regarding the orbital parameters of the
putative companion, we must recognise that the absence of a bound companion seems rather unlikely
based on the HARPS data alone: when fitting the HARPS data with a constant velocity, the reduced
$\chi^2$ with 9 degrees of freedom is equal to 3.6, and the associated probability that the model
represents the data is of only 0.02\%. Nevertheless, this result depends essentially on two HARPS
radial velocity measurements with values around 5000\,m/s in the middle of Fig.~\ref{fig:harps},
and a better time sampling would be needed to confirm the detection of the putative companion. It
must finally be noted that stellar pulsations are not expected to produce such radial velocity
variations in the case of $\zeta$\,Aql, a main-sequence A0 star not classified as a $\delta$\,Scuti
variable.

A last constraint that we can put on the potential low-mass companion is that it must comply with
the astrometric stability of $\zeta$\,Aql: {\sc Hipparcos} observed this star on about 20 different
occasions between March 1990 and March 1993 (total time span of 1000 days), and found no sign of
binarity at an astrometric accuracy of 0.72\,mas \citep{Perryman97}. Taking an estimated mass of
$0.57\,M_{\odot}$ for the companion, this basically rules out any circular binary solution with a
semi-major axis larger than 0.2\,AU at the $3\sigma$ level. However, if the orbital period is
significantly larger than the time span of the {\sc Hipparcos} observations, the presence of a
low-mass companion could have been missed, all the more that the system is supposedly viewed
edge-on so that the reflex motion of the primary star could have been confused with its proper
motion. We estimate that binary solutions with orbital periods larger than about 3000\,days, even
though they are associated with astrometric signals larger than 50\,mas, would not have been
detected at the $3\sigma$ level by {\sc Hipparcos}. This corresponds to a semi-major axis of about
5.5\,AU. All solutions with semi-major axes between 0.2\,AU and 5.5\,AU would thus most likely have
been detected by {\sc Hipparcos}, and the two most probable solutions provided by the fit to the
HARPS data seem incompatible with the astrometric constraints. However, we note that there remains
a range of plausible solutions at semi-major axes between 5.5\,AU and 8\,AU, with stellar masses
ranging between $0.6$ and $0.65\,M_{\odot}$. We also note that the use of eccentric orbits to fit
the HARPS data could lead to significantly different orbital solutions, which may improve the
compatibility with the {\sc Hipparcos} constraints.

In conclusion, the presence of a bound low-mass companion can neither be confirmed nor rejected
based on currently available data. We plan to obtain further observations of $\zeta$\,Aql at high
angular resolution with single-pupil telescopes to give a final answer to this question. In
particular, the new Sparse Aperture Masking mode of the NACO adaptive optics instrument at the VLT
would be perfectly suited to reach this goal. Interferometric closure phase measurements on longer
baselines could also help unveiling the nature of this infrared excess.



\section{Conclusion}

In this paper, we have investigated the close neighbourhood ($<1 \arcsec$) of six nearby A- and
early F-type MS stars, in search for hot counterparts to the cold debris discs detected by mid- and
far-infrared spectro-photometric space-based observations. The high-accuracy squared visibilities
collected with the CHARA/FLUOR interferometer, combined with semi-empirical models of stellar
photospheres including rotational distortion, has allowed us to reach dynamic ranges ranging from
1:175 to 1:625 at $1\sigma$ around the target stars. At this level of precision, the presence of a
resolved $K$ band emission has been identified around only $\zeta$\,Aql, with a estimated $K$-band
excess of $1.69\pm 0.31$\%. This detection adds to our previous results on $\alpha$\,Lyr
\citep{Absil06} and $\tau$\,Cet \citepalias{DiFolco07}, giving an overall near-infrared excess
detection rate of $3/9$ for the MS stars surveyed so far, among which $2/7$ are early-type stars.
The healthy statistical behaviour of the five non-detections in the present sample and the
confirmation of the excess emission around $\alpha$\,Lyr with an improved photospheric model and a
new version of the FLUOR Data Reduction Software demonstrate the robustness of our approach for hot
debris disc detection.

Our near-infrared interferometric measurements are not sampling the Fourier frequency plane in a
sufficiently dense manner to derive the morphology of the excess emission source. In particular,
both a point-like source and an extended circumstellar emission can reproduce our observations.
While in the cases of $\alpha$\,Lyr and $\tau$\,Cet, the presence of a bound or unbound companion
to the target stars within the small FLUOR field-of-view could be rejected with a high confidence,
we cannot rule out the presence of a low-mass companion in the close vicinity of $\zeta$\,Aql to
reproduce the measured $K$-band excess. The combination of the astrometric stability of
$\zeta$\,Aql measured by {\sc Hipparcos}, the variability of the radial velocities measured with
HARPS and the absence of off-axis companion in PUEO observations restricts the parameter space of
the high-contrast binary scenario to masses in the range $0.6$ to $0.65\,M_{\odot}$ and semi-major
axes between 5.5 and 8\,AU (i.e., about 200 and 300\,mas). The $K$-band contrast between the
primary and its companion would then be $\Delta K \sim 4$, making it one of the closest
high-contrast companions resolved around MS stars so far.

Besides a low-mass companion, the presence of hot circumstellar dust grains producing a significant
thermal emission in the $K$ band is another viable explanation of the observed excess emission. In
particular, we show that the debris disc model that \citet{Absil06} have proposed in the context of
$\alpha$\,Lyr is consistent with both our $K$-band detection and archival near- and mid-infrared
spectro-photometric measurements. However, our re-interpretation of archival Spitzer/MIPS and
Spitzer/IRS data clearly shows that the presence of an outer debris disc, suggested by
\citet{Chen05}, can be firmly ruled out at the sensitivity level of MIPS and IRS. In the absence of
significant amounts of cold dust, the hot debris disc scenario is not favoured to explain our
CHARA/FLUOR measurements.

The statistics of the ``hot debris disc'' phenomenon presently remains poorly constrained: for
early-type stars, a hot debris disc was found around only one star ($\alpha$\,Lyr) out of six bona
fide debris disc stars observed so far. The case of $\alpha$\,Lyr could therefore be rather
unusual. Our interferometric survey of debris disc stars will be extended to larger sample in the
coming years to improve our statistics. The case of $\zeta$\,Aql must be considered separately,
since we show that it is not surrounded by cold dust at the sensitivity level of the Spitzer
instruments, and since a close companion is a likely explanation to the observed $K$-band excess.
This star will deserve special attention in the future, and further observations will be performed
to determine the actual nature of the near-infrared excess emission that we have resolved. In
particular, infrared aperture masking experiments on large telescopes have the potential to reveal
the true nature of the observed excess.


\begin{acknowledgements}
We thank P.~J.~Goldfinger and Ch.~Farrington for their invaluable assistance with the operation of
the CHARA Array, as well as the anonymous referee for a precious help in improving the quality and
readability of the paper. O.A.\ acknowledges the financial support from the European Commission's
Sixth Framework Program as a Marie Curie Intra-European Fellow (EIF). Research at the CHARA Array
is funded by the National Science Foundation through NSF grant AST-0606958 and by the Georgia State
University College of Arts and Sciences. This research has made use of NASA's Astrophysics Data
System and of the SIMBAD database, operated at CDS (Strasbourg, France). Part of this work was
performed in the context of the ISSI team ``Exozodiacal Dust Disks and {\sc Darwin}''.
\end{acknowledgements}

\bibliographystyle{aa} 
\bibliography{Astars} 

\end{document}